\PassOptionsToPackage{table}{xcolor}
\documentclass[sigconf]{acmart}
\usepackage{amsmath,amsfonts}
\usepackage{algorithmic}
\usepackage{graphicx}
\usepackage{textcomp}
\usepackage{xcolor}
\usepackage{listings}
\usepackage{hyperref}
\usepackage{booktabs}
\usepackage{pdflscape}
\usepackage{multirow}
\usepackage{multicol,varwidth}
\usepackage{wasysym}
\usepackage{pifont}
\usepackage{csquotes}
\usepackage{makecell}
 
\usepackage[colorinlistoftodos,textwidth=1.22cm]{todonotes}

\newcommand{\raidforums}{\texttt{Forum 1}}
\newcommand{\crdclub}{\texttt{Forum 2}}
\newcommand{\xss}{\texttt{Forum 3}}
\newcommand{\fuckav}{\texttt{Forum 4}}
\newcommand{\verified}{\texttt{Forum 5}}
\newcommand{\hackforums}{\texttt{Forum 6}}
\newcommand{\omerta}{\texttt{Forum 7}}
\newcommand{\enclave}{\texttt{Forum 8}}
\newcommand{\exploit}{\texttt{Forum 9}}
\newcommand{\deeptor}{\texttt{Forum 10}}
\newcommand{\blackhatworld}{\texttt{Forum 11}}
\newcommand{\torum}{\texttt{Forum 12}}
\newcommand{\antichat}{\texttt{Forum 13}}
\newcommand{\nulled}{\texttt{Forum 14}}
\newcommand{\crackingpro}{\texttt{Forum 15}}
\newcommand{\cryptbb}{\texttt{Forum 16}}
\newcommand{\altenen}{\texttt{Forum 17}}
\newcommand{\rutor}{\texttt{Forum 18}}
\newcommand{\cebulka}{\texttt{Forum 19}}
\newcommand{\darknetcity}{\texttt{Forum 20}}
\newcommand{\reverse}{\texttt{Forum 21}}
\newcommand{\bhf}{\texttt{Forum 22}}
\newcommand{\cracked}{\texttt{Forum 23}}

\newcommand{\fyes}{\cellcolor[HTML]{9AFF99}\checkmark}
\newcommand{\fno}{\cellcolor[HTML]{FFCCC9}-}
\newcommand{\funkn}{\cellcolor[HTML]{FFFFC7}?}
\newcommand{\xmark}{\ding{55}}%

\usepackage{etoolbox}
\makeatletter
\patchcmd{\@makecaption}
  {\scshape}
  {}
  {}
  {}
\patchcmd{\@makecaption}
  {\\}
  {.\ }
  {}
  {}
\newcommand{\linebreakand}{%
  \end{@IEEEauthorhalign}
  \hfill\mbox{}\par
  \mbox{}\hfill\begin{@IEEEauthorhalign}
}
\makeatother

\def\BibTeX{{\rm B\kern-.05em{\sc i\kern-.025em b}\kern-.08em
    T\kern-.1667em\lower.7ex\hbox{E}\kern-.125emX}}

\newcommand{\feature}[1]{\textbf{\texttt{#1}}}

\acmConference[WEIS '23]{The 22nd Workshop on the Economics of Information Security}{July 05--08,
  2023}{Geneva, Switzerland}

\setcopyright{none}

\renewcommand\footnotetextcopyrightpermission[1]{} 

\begin{document}
\title{You Can Tell a Cybercriminal by the Company they Keep: \\A Framework to Infer the Relevance of Underground Communities to the Threat Landscape}

\author{Michele Campobasso}
\email{m.campobasso@tue.nl}
\affiliation{
\country{}
}    

\author{Radu R\u{a}dulescu}
\email{r.radulescu@student.tue.nl}
\affiliation{
\country{}
}

\author{Sylvan Brons}
\email{s.h.j.p.brons@student.tue.nl}
\affiliation{
\country{}
}

\author{Luca Allodi}
\email{l.allodi@tue.nl}
\affiliation{%
  \institution{\hspace{-\columnwidth/2} Eindhoven University of Technology}
  \country{\hspace{-\columnwidth/2} Eindhoven, The Netherlands}}

\begin{abstract}
The criminal underground is populated with forum marketplaces where, allegedly, cybercriminals share and trade knowledge, skills, and cybercrime products. 
However, it is still unclear whether all marketplaces matter the same in the overall threat landscape. 
To effectively support trade and avoid degenerating into scams-for-scammers places, underground markets must address fundamental economic problems (such as moral hazard, adverse selection) that enable the exchange of actual technology and cybercrime products (as opposed to repackaged malware or years-old password databases). From the relevant literature and manual investigation, we identify several mechanisms that marketplaces implement to mitigate these problems, and we condense them into a market evaluation framework based on the Business Model Canvas. We use this framework to evaluate which mechanisms `successful' marketplaces have in place, and whether these differ from those employed by `unsuccessful' marketplaces. We test the framework on 23 underground forum markets by searching 836 aliases of indicted cybercriminals to identify `successful' marketplaces. We find evidence that marketplaces whose administrators are impartial in trade, verify their sellers, and have the right economic incentives to keep the market functional are more likely to be credible sources of threat.
\end{abstract}

\maketitle


\section{Introduction}
\label{sec:intro}

The cybercrime underground is the subject of numerous scientific studies, and it is the source of a conspicuous volume of threat intelligence and threat information employed for a large set of objectives, including situational awareness~\cite{campobasso2020impersonation, pastrana2019measuring}, criminal behavior~\cite{gambetta2011codes, soudijn2012cybercrime, huang2018systematically}, informing law enforcement actions and intervention~\cite{christin2013traveling, thomas2015framing, huang2018systematically, alrwais2017under}, and instrumenting security countermeasures~\cite{campobasso2020impersonation, huang2018systematically}.
To the best of our knowledge, at least part of these applications implicitly assume that if cybercrime-related activities happen in a specific forum, then they must matter to the overall threat landscape. 
Yet, previous work already questioned whether this really is the case~\cite{herley2010nobody}: enabling effective trade in the underground communities is a hard, foundational problem that requires the right incentives and control mechanisms to be in place for `good sellers and buyers' (as opposed to scammers) to have an opportunity to operate in the market in the first place. In fact, like in any other market, cybercriminals capable of generating actual value seek the opportunity to appropriately price their products and services, and to clearly differentiate themselves from their inept (or plainly fraudulent) competitors~\cite{akerlof}. 
If `good sellers and buyers' (i.e., those that can supply and consume effective, real attack technology and cybercrime products) are pushed out of the market, the remaining activity, made from mostly wanna-be criminals scamming each other with repackaged old malware or `0-day exploits' downloaded from \texttt{exploit-db}, will hardly pose any plausible threat.
We thus argue that the presence of appropriate mechanisms promoting and supporting the trade of technologies and services with high intrinsic value is a key element of a credible underground market. 

In this paper, we develop an evaluation framework identifying market features addressing \textit{moral hazard} and \textit{adverse selection} issues in underground markets. We then evaluate the composition of these features across a set of $23$ underground markets to assess whether differences emerge between `successful' and `unsuccessful' markets. As no clear and objective classification currently exists of what a `successful' market is, we consider whether we can find evidence that trade activity happened in that market in the form of advertisements or interactions from sellers that led to the arrest by law enforcement of at least a cybercriminal for selling criminal products or services. The rationale for this decision is that a law enforcement arrest is a testament to the real-world relevance of the specific crime for which the warrant was issued. Therefore, we consider finding evidence linking that specific crime to trade in a specific forum as evidence that that forum is at least in principle capable of supporting the trade of effective criminal technologies, data, or services. Obviously, \textit{not} finding evidence of that trade does not mean that the market cannot support it. On the other hand, it does highlight the choice of the arrested criminals to trade in a set of markets, but not others\footnote{In this paper, we will call these markets `successful' or `unsuccessful' for simplicity. However, the reader should keep in mind these considerations when interpreting these terms.}.
Therefore, we can infer whether convicted criminals providing effective attack technology or services were more likely to sell their goods in market forums with certain features than others. Following this criterion, we find that, in general, successful markets tend to be more similar to each other, over the feature set we define in our framework than to other forum markets. Interestingly, we also find this to be true within groups: that is, successful forum markets are more similar to each other than unsuccessful forum markets are among each other, suggesting that specific features should be present to support effective trade. We proceed with investigating these features in detail and find that, among others, the lack of involvement in trade from market operators in the market they administer plays a positive role in the odds of identifying a convicted criminal in the same market. In the discussion, we offer an interpretation of the implications that these features have, and how these could identify different business models for the markets, in accordance with the goals of the market administrators. 

Section~\ref{sec:background} provides a background on the hurdles of conducting trade of criminal goods in marketplaces. Section~\ref{sec:methodology} describes the methodology used to select a set of underground communities, label them according to the framework and then assess whether a community is successful or not. Section~\ref{sec:framework} offers an overview of the framework used to classify communities based on the derived set of features. Section~\ref{sec:results} presents the results of the applied methodology, in particular the results of the validation process of the framework, and provides a qualitative analysis of the gathered data, and Section~\ref{sec:discussion} provides an interpretation of the results. Section~\ref{sec:limitations} offers a thorough overview of the limitations of this study. Finally, Section~\ref{sec:relatedwork} discusses the relevant related work, and Section~\ref{sec:conclusions} concludes the paper.


\section{Background}
\label{sec:background}
Cybercriminals meet online in forums to trade the byproducts of their attacks, such as leaks and initial access, stolen credentials and credit card numbers, as well as the attack technology itself and to exchange knowledge and meet new partners~\cite{yip2013forums, allodi2016then, thomas2015framing, soudijn2012cybercrime}. Dozens of communities are scattered across the `surface' and `deep' web, and the majority of those are easy to find and get access to. However, prior research suggests a potentially large portion of these markets may not support the trade of valuable products, featuring obsolete malware, expired credit cards, old dumps of leaked passwords, and generally publicly available information repackaged as `hacking tools' or `password leaks'~\cite{dupont2016ecology, dupont2017darkode, herley2010nobody, soudijn2012cybercrime}. 
Arguably, the reason for the unconvincing quality of their offer is an effect of the lack of mechanisms that establish trust and regulate trade, which have a key role in any transaction, let alone those between mutually distrustful parties~\cite[Chap.2]{gambetta2011codes}, \cite{motoyama2011analysis, allodi2016then, dupont2017darkode, yip2013forums, yip2013trust, soudijn2012cybercrime}. As initially underlined by Herley and Flor{\^e}ncio~\cite{herley2010nobody}, in the absence of trade regulation, these markets assume the typical characteristics of `markets for lemons'~\cite{akerlof}: these markets are fraught with quality uncertainty and operate under constraints (and related risks) of strong information asymmetry~\cite{allodi2016then, yip2013trust, herley2010nobody, allodi2017economic, soudijn2012cybercrime}.
These issues materialize both at the level of market participants, as well as the market administration level, and each identifies different problem dimensions.

\subsection{Market participation}
The \textit{principal-agent problem} captures the dynamics between two parties (namely, the \textit{agent}, who is the person who acts on behalf of someone else, the \textit{principal}), where each party acts both in function of an agreement (contract) with their counterpart and in the pursuit of self-interests~\cite{gailmard2014principal, eisenhardt1989agency}. 
In general, Agency Theory distinguishes \textit{ex-ante} from \textit{ex-post} risks, depending on whether they materialize before or after the stipulated contract between agent and principal is enforced~\cite{chohan2020opportunistic}. Among ex-ante risks, \textit{adverse selection} affects the ability of a buyer to correctly choose the service or product they need. Among ex-post risks, \textit{moral hazard} risks materialize if either party in the contract unilaterally changes their behavior, \textit{de facto} exploiting some weaknesses of the contract, or outright dismissing their obligations toward the other party. This risk is generally present when one of the parties does not bear the full costs of their actions, i.e., there is no expectation of punishment or penalty (`shadow of the future') for the misbehavior.

\smallskip 
\noindent
\textbf{Adverse selection.} 
\textit{Information asymmetry} is a common issue in cybercriminal markets; the problem is exacerbated by the (often) impossibility of sellers to provide detailed information about the quality of their products, leading to \textit{adverse selection} (i.e., buyers can not distinguish between products of different quality, or to choose the correct one for their needs). In fact, providing details on the product's effectiveness from the seller (e.g., how a 0-day exploit works, validity of credit cards) could spoil the value of the product itself by revealing critical information to reproduce the artefact without purchasing it. In other words, it is difficult for a buyer to assess the quality of an offered product, and buyers can often fall victim to scams. That creates an incentive for scammers, which offer their low-quality products (or even no product at all) for competitive prices. As Akerlof noted in his seminal work on the `market for lemons'~\cite{akerlof}, since buyers have no way to evaluate the qualities of the offered products, these markets suffer two major pitfalls: first, cheap (and low-quality) products become indistinguishable from better (and therefore likely more expensive) products; this pushes `honest' sellers out of the market as their profit margins quickly become unsustainable (or negative) as the `competition' can set prices arbitrarily low. Second, the exit of `honest' sellers from the market spreads awareness among market participants that all products are of bad quality, leading to a failed market. 

\smallskip 
\noindent
\textbf{Moral hazard.}
Cybercrime benefits from a lower risk of exposure and arrests thanks to the use of privacy-enhancing technologies and the relatively slow response of law enforcement prosecution~\cite{lusthaus2012trust, dupont2016ecology}. 
However, this creates a challenge for cybercrooks to correctly identify their fellow criminals. In fact, in the lack of social control and signaling mechanisms typical of traditional crime~\cite{dupont2016ecology, lusthaus2012trust, gambetta2011codes} (which may resort to the use of violence to ensure contract compliance~\cite{pearson2003king}), establishing and maintaining trust relationships with co-offenders becomes a hurdle. As an effect, dishonest cybercriminals could see the opportunity to easily join communities to advertise products that will never be delivered, without exposing their identity. When scammers have little to no fear of punishment, potential victims may incur \textit{moral hazard}. Moral hazard is the risk that one of the two parties during a transaction violates the agreed terms during its execution (e.g., the product is not delivered after payment). 

\subsection{Market administration}
A different set of dimensions relate to how the market is administered, and in particular on what incentives are present to sustain the market.

\smallskip
\noindent
\textbf{Cost structure and risks.}
Maintaining an underground community with (sometimes hundreds of) thousands of members is a costly venture~\cite{goncharov2015criminal, alrwais2017under}. Furthermore, such marketplaces often are victims of DDoS attacks from competitor platforms~\cite{collier2020cybercrime}, and the nature of their content is an appealing target for law enforcement. These problems further increase costs, as it calls for the use of bulletproof hosting and DDoS mitigation services from the gray markets, which may apply higher fees than commercial solutions due to the shortage of competent providers~\cite{collier2019booting}. 
Therefore, market administrators need to find the funds to run and scale up their infrastructure as the community grows. 
Common strategies adopted by market administrators are to ask their members taxes for conducting business, running ad campaigns, and imposing registration fees~\cite{georgoulias2021qualitative}. Arguably, community contributions (e.g., in the form of account verification fees) to the maintenance of the community can be seen as a positive incentive in making fraudulent options (e.g., exit scams) less attractive. 

Further, underground communities face the problem of attracting the interest of threat intelligence companies, researchers, and law enforcement, which monitor their activity to gauge the threat levels they pose, study their functioning, aim at undermining their operations by disrupting the correct functioning of their reputation systems~\cite{herley2010nobody, franklin2007inquiry}, or perform market take-downs. In addition, competitor communities or disgruntled members may attack the market to exfiltrate private data from the forum (e.g., market `leaks')~\cite{motoyama2011analysis, overdorf2018under} and damage their reputation. To face these threats respectively CAPTCHAs and anti-DDoS services are employed to discourage attackers and limit the effectiveness of vulnerability scanners, granting some more privacy to the community members.

\smallskip \noindent
\textbf{Underground markets as business endeavours.} 
In this sense, forum cybercriminal markets can be thought of as businesses. To create a functional marketplace, administrators need to account for the needs of their members (both customers and sellers) to effectively support trade, as well as the cost structure and the operative risks they face. As we discussed before, reliably running and maintaining a marketplace comes at a cost; hence, market operators need to carefully model their business in function of two stakeholders: themselves, to obtain revenue (and profits), and market participants, to create transaction volume. The Business Model Canvas (BMC) introduced by Osterwalder~\cite{osterwalder2010business} allows for an overview of an existing business model or supports the creation of a new one. The BMC could be used to decompose the value chain of the market in its key components, to understand how the value added from the market is supported by a careful selection of their suppliers and activities that produce value to customers that need to be reached and taken care of while making a profit. Thus, exploring underground communities from this angle could potentially indicate flaws in the business model of some markets, indicating structural problems that must be accounted for when a new criminal community is established. 


\section{Approach} 
\label{sec:methodology}

In this section, we detail our approach to the definition of the framework and its evaluation.

\subsection{Selection of underground communities}
\label{subsec:methodology_communities}

The selection of communities involved in this study is the result of two years of market infiltration and monitoring. Forum markets are included in our selection opportunistically, with network effects (e.g., interactions within the communities pointing at other forums not yet on our list) accounting for the majority of the market identification mechanism. The resulting list includes markets oftentimes studied in the scientific literature, such as \raidforums, \hackforums, \cracked, \antichat, \blackhatworld, and \nulled~\cite{bermudez2021shady, pastrana2019measuring, bhalerao2019mapping, afroz2013honor}, as well as less frequently studied (to the best of our knowledge) markets such as \exploit, \verified, \xss, and \torum.

Overall, this study involves the analysis of $23$ underground communities, $19$ of which are still active and $4$ of which are deceased for which a data collection is available; of these, $3$ are obtained from data shared by Web-IQ, an industry player active in the domain of cyber threat intelligence sharing; the fourth is a market we had previous access to for which we have a dump. 
Table~\ref{tab:communities}
\begin{table*}
    \centering
    \scalebox{0.89}{
    \begin{tabular}{l|lllp{0.4\textwidth}lll}
    \toprule
        \textbf{Market name}     & \textbf{Members} & \textbf{Threads}   & \textbf{Posts}         & \textbf{Description} & \textbf{Segregated} & \textbf{Alive} & \textbf{First activity} \\
        \midrule
        \raidforums$^1$ & $748'348$     & $121'271$ & $3'821'014$   & Leaked databases, hosting, proxies, combolists, stolen accounts & \xmark & \xmark & Apr 2015$^\dagger$ \\
        \crdclub       & $224'213$ & $97'308$ & $429'608$ & Carding, counterfeit documents, traffic & \xmark & \checkmark & Jun 2013$^\ddagger$ \\
        \xss & $52'813$ & $62'170$ & $484'452$ & IAB, Exploits, malware, malware development, spam, carding, traffic, hosting, counterfeit documents, dating scams & \xmark & \checkmark & May 2005$^\dagger$ \\
        \fuckav$^2$      & $43'638$ & $29'137$ & $119'874$ & Malware development, obfuscation, and AV evasion, hosting, proxies  & \xmark & \xmark & Aug 2009$^\dagger$ \\
        \verified        & $117'850$ & $264'564$ & $1'246'883$ & Wire fraud, cashing out, carding, malware, botnets, hosting, dating scams & \checkmark & \checkmark & May 2005 \\
        \hackforums      & $5'326'374$ & $6'234'742$ & $61'891'796$ & Malware, botnets, stolen accounts, gaming hacks, fraud schemes &  \xmark & \checkmark & May 2007$^\dagger$ \\
        \omerta     & $85'303$ & $46'101$ & $376'396$ & Carding, counterfeit documents, stolen accounts & \checkmark & \checkmark & Dec 2010 \\
        \enclave         & $19'772$ & $4'173$ & $46'288$ & Carding, hosting, cashing out, stolen accounts & \xmark & \checkmark & Sep 2016 \\
        \exploit         & $81'558$ & $201'412$ & $1'275'035$ & Exploits, malware, malware development, IAB, hosting, spam, traffic, financial fraud & \checkmark & \checkmark & Feb 2005 \\
        \deeptor         & $63'711$ & $12'880$ & $140'440$ & Carding, stolen accounts, database leaks, porn, financial fraud & \xmark & \checkmark & Jul 2015 \\
        \blackhatworld   & $1'717'605$ & $1'479'148$ & $16'155'076$ & Traffic, hosting, financial fraud, malvertisement, proxies, SEO fraud, PPI & \xmark & \checkmark & Oct 2005 \\
        \torum$^3$           & $54'829$ & ? & $71'466$ & Carding, malware & \xmark & \xmark & May 2017\textsuperscript{$\dagger\dagger$} \\
        \antichat        & $147'809$ & $201'189$ & $2'432'625$ & SEO fraud, proxies, hosting, traffic, malware development & \xmark & \checkmark & Aug 2001$^\ddagger$ \\
        \nulled          & $4'801'334$ & $1'273'687$ & $38'118'954$ & Combolists, proxies, stolen accounts, leaked databases, financial fraud, dating scams & \xmark & \checkmark & Jan 2015 \\
        \crackingpro     & $333'147$ & $186'458$ & $2'028'241$ & Stolen accounts, porn, proxies, combolists & \xmark & \checkmark & Dec 2016 \\
        \cryptbb         & $65'911$ & $12'664$ & $40'877$ & Malware, financial fraud, hosting & \checkmark & \checkmark & Jul 2017 \\
        \altenen         & $1'245'284$ & $1'117'023$ & $8'086'074$ & Financial fraud, porn, carding, proxies & \xmark & \checkmark & May 2018$^\ddagger$  \\
        \rutor           & $320'964$ & $39'722$ & $4'360'559$ & Cashing out, drugs, financial fraud, information gathering, weapons, counterfeit documents, money laundering & \xmark & \checkmark & Oct 2014 \\
        \cebulka         & $45'262$ & $17'017$ & $63'937$ & Data leaks, drugs, banking accounts, counterfeit documents, weapons & \xmark & \checkmark & Feb 2016 \\
        \darknetcity     & ? & ? & ? & Carding, travel fraud, hosting, financial fraud & \checkmark & \xmark & Jan 2021 (?)\textsuperscript{$\dagger\dagger$} \\
        \reverse         & $7'205$ & $1'449$ & $28'119$ & reverse engineering, hacking tools, game hacks, malware, malware obfuscation & \xmark & \checkmark & Apr 2010$^\dagger$ \\
        \bhf             & $311'015$ & $555'208$ & $5'641'640$ & DDoS, proxy, hosting, game items, carding, PPI, malware, leaks & \xmark & \checkmark &  Nov 2016 \\
        \cracked         & $3'594'708$ & $769'668$ & $23'176'695$ & Financial fraud, porn, dating scams, stolen accounts, counterfeit documents, proxies & \xmark & \checkmark & Mar 2018 \\
        \bottomrule
    \end{tabular}}
    \vspace{1mm}
    \caption{Overview of analyzed forum communities}
    \begin{minipage}{0.93\textwidth}
    \footnotesize All the reported figures have been fetched in Mar 1\textsuperscript{st}, 2023, or otherwise specified with the latest information available; $^1$: Feb 23\textsuperscript{rd}, 2022, due to market closure (source \verb|archive.org|); $^2$: Aug 31\textsuperscript{st}, 2021, due to market closure (source Web-IQ); $^3$: Dec 21\textsuperscript{st}, 2019, due to market closure (source Web-IQ). In the column first activity, we report the registration date of forum administrators, unless otherwise specified; $^\dagger$: oldest \verb|archive.org| copy available; $^\ddagger$: whois domain registration date; \textsuperscript{$\dagger\dagger$}: first available information found online 
    \end{minipage}
    \label{tab:communities}
\end{table*}
provides an overview of the forum communities included in this study. 
A community is indicated as `alive' if at the date of the study (Feb 2023) the community is active and reachable on the Internet. We report as `segregated' communities that require vouching, screening, or upfront payment before access to the community is granted. 
Among these, we had long-standing access to \exploit, via invitation. To gain access to \cryptbb\ we were screened for knowledge of penetration testing and for our motivation to join the community (for which we gave generic answers indicating interest in potential commercial opportunities). We registered to \darknetcity\ (currently a deceased market) during a period when free registration to the market was available. To access markets imposing a paywall at registration (\verified\ and \omerta), we paid the registration fee\footnote{We discuss the ethical implications in Section~\ref{subsec:ethical}}. We evaluated the benefit of accessing these communities based on the overall feedback and `reputation' those had within the communities we already had access to. 
We purchased premium subscriptions for five communities to get full access to their content, namely \deeptor, \crackingpro, \nulled, \altenen, and \cracked. 
To collect data from the markets, we developed four simple crawlers, configured to interact with the $20$ ($19$ still alive markets now, plus \darknetcity\ which at the time of our first investigation was still active) marketplaces to identify users of interest (see Section~\ref{subsec:methodology_validation}) and related activity. To minimize crawler exposure during the activity, we follow~\cite{campobasso2022threat} and employ the browser instrumentation library \texttt{Selenium}~\cite{selenium} and \texttt{tbselenium}~\cite{tbselenium}.
We provide a discussion on ethical considerations for the data collection in Section~\ref{subsec:ethical}. 

\subsection{Framework derivation and instantiation}
\label{subsec:methodology_framework_derivation}

We structure our framework over the dimensions identified in Section~\ref{sec:background} (\textit{Market participants:} moral hazard, adverse selection; \textit{Market administration:} cost structure and risks). 
To finalize our framework, we first identify from the literature mechanisms commonly employed by underground forum markets that map to the identified dimensions (e.g., a reputation system is a mechanism addressing both adverse selection and moral hazard). To identify the relevant literature, we queried Google Scholar using combinations of the following keywords: \texttt{cybercrime}, \texttt{underground}, \texttt{forums}, \texttt{markets}, \texttt{moral hazard}, \texttt{adverse selection}, \texttt{trust}. We limited our research to the top 5 pages of results for each query. We read the papers and included those discussing the operational and economic factors of underground markets, and analyzed the relevant related works of each paper. This led us to identify 23 studies discussing underground market forums characteristics to different degrees. We report the identified mechanisms and map them to the literature in Section~\ref{sec:mechanisms}.
Table~\ref{tab:theorymapping} provides an overview of the mapping. 

However, these `dimensions' are generally high-level features that can be implemented differently and with different strategies by each forum (e.g., some markets may restrict reputation changes to happen only after trade, and others may have no restrictions on who can assign `reputation points' to a user). To capture these differences, we adopt a two-iterations `bottom-up' approach whereby for each market feature we first identify within the set of $23$ markets under analysis what concrete features are implemented in support of the identified mechanisms, and refine the framework accordingly. During this procedure, we pay particular attention to identifying mechanisms adopted by different marketplaces, and evaluate them in relation to the problem dimensions (Section~\ref{sec:background}) for inclusion in the framework. 

Once all features enumerated from the market observations are included in the framework, we re-iterate across all $23$ market forums to evaluate whether a certain feature is present or not in any given market. We follow this process (rather than assigning features to markets as we go by in the first iteration) as some features may only be `implicitly' present in a market, and be ignored at the first pass. This way we assure that all marketplaces are evaluated over the same set of features. Each feature can be labeled as \texttt{present}, \texttt{non-present}, or \texttt{unknown}. 

To aid the analysis of the identified features from the perspective of the market's `business proposition', we further map the identified features to the BMC~\cite{osterwalder2010business}. 
For details on the mapping, we refer the reader to Figure~\ref{fig:bmc} and associated discussion. 
We report on all the identified features, and their mappings, in Section~\ref{sec:framework}. 

\subsection{Framework validation}
\label{subsec:methodology_validation}

\noindent\textbf{Ground truth definition.} To evaluate whether the presence of the identified market features correlates with or signals the ability of a market to support successful trade of effective criminal technology, we first need to build a ground truth of which markets in our collection can be considered `successful'. As there is no commonly accepted definition of a `successful market', we define it as a market for which there is evidence that it is capable of supporting the trade of real, effective criminal technology or services. To do this we want to identify classification criteria that are (a) objective; (b) independent from our analysis; (c) unambiguous; (d) credibly measurable in our data.
Following these criteria, we consider whether (at least one) convicted cybercriminal relied on a specific market to sell technology or services for which they were arrested and convicted. 
This of course limits our definition of success to those markets whose users have been eventually convicted, and risks overlooking `successful' (perhaps, `even more successful') markets no member of which has been arrested yet (or for which an arrest cannot be linked to that market). On the other hand, it does speak about what choices those convicted cybercriminals made when deciding in which market to trade their goods. Very importantly, this also \textit{explicitly defines the scope within which the claims of this study should be interpreted}.
The adopted criterion is (a) objective, because it considers the real-world impact of the crimes committed and enabled through that marketplace; (b) independent from our analysis, as the investigation leading to a warrant, arrest, and ultimately the conviction of a cybercriminal are run by law enforcement and played no role in the forum selection considered for this study; (c) unambiguous, because the employment of a technology or service leading to a conviction is clearly working and effective; (d) credibly measurable in the data, because trial and warrant documents come rich in information about the reasons for the arrest and any indication of the online presence of the criminal, which can be at least partially mapped back to the criminal's activity in the underground market space. Details on how we process this information follow. 

\smallskip
\noindent\textbf{Ground truth measurement.} To obtain a list of convicted cybercriminals, we rely on the \verb|arresttracker.com|~\cite{arresttracker} cybercrime online database\footnote{As of today, the website is down. We tried to get in touch with its administration but without success. We share the dataset at \url{https://security1.win.tue.nl}.}. The dataset contains 2775 individual cyber-related crime incidents spread from 1970 to 2021, documented from warrants and indictments from the USA Department of Justice, Europol press releases, and media outlets\footnote{We discuss the limitations of this dataset in Section~\ref{sec:limitations}}. An individual entry in the database corresponds to a convicted criminal and contains information fields about their online presence (i.e., their online \textit{alias}) and technical details on the specific crime (e.g., stealing/selling credit card data, offering specific malware or criminal services). 
For this research, we consider only criminals arrested after (and including) 2011, and for which an alias is indicated. We disregard cases before 2011 to limit the biases caused by not accounting for the restricted (and different) forum choices available to cybercriminals in 2011 ($\approx35\%$ forums created before 2011) in our sample. That resulted in 836 aliases of 480 cybercriminals indicted in the 2011-2021 interval. The data available on \texttt{arresttracker.com} terminates in December 2021 and the project is abandoned. 
We define three classes of possible matches: 
\begin{itemize}
    \item \textit{clear hit}: a forum user in any of the considered forum markets whose username corresponds exactly to an alias reported in a conviction report, \textit{and} that presents activity in that forum that closely resembles the crime for which the related criminal has been arrested and convicted. For example, if a username in a forum corresponds to the alias of a criminal convicted for running and renting a botnet, we only consider a \textit{clear hit} if the forum activity of that username is directly related to running and renting a botnet. Furthermore, we check whether the timelines of a user activity on a forum and a conviction are compatible. This is to avoid misclassifying possible copycats.
    \item \textit{inconclusive hit}: a forum user whose username exactly matches the alias of a convicted criminal but whose activity on the forum (whereas related to cybercrime, e.g., selling stolen data) does not match the reason for the arrest (e.g., running and renting a botnet). Furthermore, we include those cases where the activity on the forum matches the reason for the arrest, but there is no convincing evidence of trade.
    \item \textit{no hit}: a forum username matches an alias, but no activity related to cybercrime is found.
\end{itemize}

Each market may have none, only one, or multiple matches for each class.
We consider `clear hits' as the strongest evidence that a convicted criminal chose a specific forum market (or multiple ones) to trade effective criminal technology. 

\smallskip
\noindent\textbf{Example of `hits'.} To exemplify, we consider the case of Alexey Klimenko, accused of being part of the Infraud Organization as the provider of bulletproof hosting to create, operate, maintain, and protect their own online contraband stores~\cite{klimenko2}. According to the superseding indictment, dated Oct 31\textsuperscript{st}, 2017, Alexey Klimenko was operating under the alias of `Grandhost'. We found account usernames that match the alias `Grandhost' on six marketplaces in our data: \fuckav, \exploit, \omerta, \verified, \xss, and \antichat; for \omerta, the registered user exists but has no messages nor activity in the forum. Therefore, we classify this alias in that market as a `no hit'. In \verified\ and \antichat, we find evidence that the user `Grandhost' interacted with members of the community on topics related to bulletproof hosting, but we could not find any evidence of trade of services or products on those forums linked to bulletproof hosting for this user. Therefore, we classify the occurrences of `Grandhost' in \verified\ and \antichat\ as `inconclusive hits'. On the three remaining marketplaces (\fuckav, \exploit\ and \xss) we found three threads created by the user `Grandhost' titled (with little variations of) `Bulletproof Hosting Service' (translated from Russian). In those, the author advertises services related to dedicated servers, VPS/VDS, bulletproof domains, and SSL certification, providing a range of locations for their servers, and exemplifying some of the possible (unlawful) uses they allow. The first advertisements date February 2010 in both \exploit\ and \xss, and a similar one follows the same year in July on \fuckav. Up until February 2016, on all three markets, product advertisement is updated by `Grandhost' when new offers or features are available to customers; this suggests a long-standing commercial activity for that user in all three forums. The last recorded activity in our data for the user `Grandhost' is in June 2016, on \exploit. The most recent evidence of trade for Grandhost's offerings we find on \exploit\ in the form of user (positive) feedback left in the forum on July 2017. Currently, the account for `Grandhost' on \exploit\ appears as `deactivated'. We note that the timeline of the forum activity is also compatible with the timeline of the conviction. Hence, we classify the alias `Grandhost' as a `clear hit' for \fuckav, \exploit, and \xss.

We repeat this process for all username-alias matches we find in our data. Section~\ref{subsec:results_aliases} provides an overview of results. 

\smallskip
\noindent\textbf{Evaluation.} To evaluate whether the identified feature set correlates to the ground truth, we compare whether (a) `successful' forums are more similar to each other over the defined feature set than they are to `unsuccessful' forums; this test allows us to evaluate whether specific feature compositions are proper of `successful' markets but not of `unsuccessful' markets. Further, we test whether (b) `successful' forums are more similar to each other than they are with `unsuccessful' ones; this test allows us to evaluate whether the feature set of `successful' forums tends to be more stable than the feature set of `unsuccessful' forums. To evaluate similarity (for both tests), we compute the Jaccard distance for all forum pairings and employ two one-sided Wilcoxon rank sum tests to evaluate the hypotheses above.

Further, we explore whether we can find differences at the level of specific groups of features (as defined by the BMC), as well as at the level of the specific features between the two groups of forums. To do this, we constrain all features we identify to have the same `direction' (i.e., all should contribute positively to the mitigation of either adverse selection or moral hazard). Under this constraint, we perform a set of Fisher exact tests evaluating the count of forums of each type (`successful', `unsuccessful') for which the tested set of features is present.

For all tests, we consider a statistical significance level $\alpha=0.1$ as the threshold for significance. We note that, given the inherently qualitative nature of this study, we consider statistical tests as merely a guide to drive the discussion around our findings, rather than a quantitative assessment of their statistical robustness.

\subsection{Ethical aspects}
\label{subsec:ethical}

To conduct this study, we had to gain complete access to a number of underground communities and crawl them, which required us to infiltrate, sometimes interact with community members, or even pay a fee for registering or to get full access. 

\smallskip
\noindent \textbf{Crawling.} Our crawling activity was performed in a way that: (1) our crawlers were solely verifying the existence of usernames in the forums via the appropriate search function, (2) we relied on the Tor network only when the target community was only reachable via it, (3) we limited the bandwidth usage of the crawler to be more similar to a human when fetching new content (especially when the use of Tor was necessary), in accordance to the considerations of~\cite{campobasso2022threat}, helping us maintaining stealth and avoiding compromising the stability of the target~\cite{christin2013traveling}. Even though collecting data from underground marketplaces could be considered an activity not in agreement with their terms of service~\cite{pastrana2018crimebb}, studies highlight that the societal benefit of studying cybercriminal ventures outweighs these concerns~\cite{martin2016ethics, benjamin2019dice}.

\smallskip
\noindent \textbf{Paying access.}  
During our investigation, we identified a number of forums requiring to pay a fee to register or to access portions of the forum. Respectively, from discussions in affiliated communities or within the same communities, we made considerations on whether the benefit from obtaining access could justify the expense. As pointed out by Benjamin~\cite{benjamin2019dice}, when the information that could be made accessible could be of great value for the research (in our case, to avoid undermining the internal validity of our study), this should be considered and be a matter of discussion with the relevant ERB. For paywalls at registration, we only considered two markets for which subscription fees did not exceed $100$ USD; in the case of private sections, we often had access to a 1-month premium subscription for $\approx10$ USD. In cases where fees were higher or came with additional problems (e.g., \blackhatworld\ requires a fee of $147$ USD and $100$ posts to obtain the upgrade), we refrained to proceed with the purchase. We discussed our project with our institutional ERB, and we received approval with reference number ERB2021MCS1.


\section{The framework} 
\label{sec:framework}

\subsection{Identification of market mechanisms addressing moral hazard and adverse selection}
\label{sec:mechanisms}

In this section, we identify from the literature the key mechanisms forum market (administrators) put in place and link them to foundational dimensions identified in Section~\ref{sec:background}. Table~\ref{tab:theorymapping} provides an overview of the identified mechanisms and the rationale for the mapping.

\begin{table*}[t]
\centering
\caption{Mapping market mechanisms to adverse selection, moral hazard, cost structure and risks.}
\label{tab:theorymapping}
\small
\begin{tabular}{p{0.1\textwidth}p{0.09\textwidth}p{0.09\textwidth}p{0.65\textwidth}}
\toprule
\textbf{Foundational aspect} & \textbf{Key issue} & \textbf{Market \newline mechanism} & \textbf{Mapping rationale} \\
\midrule
    Market \newline participation & Adverse \newline selection & Reputation \newline systems & The presence of a reputation system is oftentimes enabled by a feedback mechanism that allows potential buyers (and possibly sellers) to assess the nature of the other party, as well as of their products (e.g., through product reviews or open feedback from previous buyers).\\
    \cmidrule{3-4}
    & & Interaction model & The interaction between community members enables communication between buyers and sellers, helping customers in the product assessment process and sharing information with future customers (in case of public interaction). \\
\cmidrule{2-4}

    & Moral hazard & Reputation \newline systems & A reputation system may serve as a `punishment mechanism' for misbehaving users, and provide an indication of future (expected) behavior.\\
    \cmidrule{3-4}
    & & Restricted \newline access model & Limiting access to the forum market to selected users that have either been vouched by existing members and/or the forum administrators or that committed to their participation in the forum via payment of an entry fee.\\
    \cmidrule{3-4}
    & & Escrow for \newline payments & The presence of escrow services may help to establish trust after one of the two parties commits to his part of the agreement, thus serving as a fallback mechanism to avoid scams caused by, for example, undelivered or non-functional goods for which a sum of money has already been transferred (in this case, to the escrow) by the buyer.\\
    \cmidrule{3-4}
    & & Dispute \newline resolution \newline system & A dispute resolution system may allow market participants to voice their claims and seek justice for unfair behavior. Importantly, it enables punishment mechanisms (such as labeling a user as a scammer and/or banning them from the community), which may deter unfair behavior.\\
    \cmidrule{3-4}
    & & Rule system & The presence of rules on trade and punishments in case of misconduct, together with their enforcement, may act as a deterrent for misconduct. \\
\midrule

    Market \newline administration & Cost structure and risks & Market \newline business plan & Clear revenue strategies for market administrators should guarantee that they have the right incentives to continue running the market as a source of revenue, rather than exit-scamming. \\
    \cmidrule{3-4}
    & & CAPTCHAs and DDoS protection & The presence of custom-created and commercial CAPTCHAs poses a threat to the execution of a crawler, limiting the possibility of exfiltrating data from the market, while DDoS protection could avoid costly downtimes. \\
    
\bottomrule
\end{tabular}
\end{table*}

\subsubsection{Restricted access model}
\label{subsubsec:framework_access}
Different access mechanisms regulate the influx of members to a community or parts of it. Some studies show that markets present different levels of segregation, ranging from open access markets where registration only requires a valid email address~\cite{herley2010nobody, allodi2017economic, dupont2016ecology, georgoulias2021qualitative}, up to access mechanisms that require the payment of a fee and perform `background checks' on the applicant~\cite{allodi2017economic, turk2020tight, georgoulias2021qualitative, laferriere2023examining, huang2018systematically, yip2013trust, thomas2015framing}; these access mechanisms may be enforced to grant access to the forum as a whole~\cite{dupont2017darkode, lusthaus2012trust} or only to portions of it~\cite{motoyama2011analysis}. While stricter access mechanisms pose a barrier for new members, these allow screening new applicants by proving their intentions either economically (i.e., by paying a registration fee), by receiving `vouches' from members of the community who grant on the intentions or identity of the new member, or via interview. These barriers potentially discourage `dishonest' merchants from joining the community by increasing the costs associated with the planned fraudulent activities~\cite{huang2018systematically}, thus potentially creating a more selective and trustworthy community~\cite{herley2010nobody, allodi2017economic, allodi2016then, yip2013trust}.

\subsubsection{Dispute resolution system}
\label{subsubsec:framework_dispute}
The lack of accountability for misbehavior in the underground calls for strategies to take action against miscreants and to settle disputes~\cite{dupont2017darkode, lusthaus2012trust, georgoulias2021qualitative}.
Dispute resolution systems aim at settling a disagreement between two (or more) community users, generally caused by one of the two parties not adhering to the conditions agreed upon in trade. Typically, the offended part of the claim (the plaintiff) can open a request for arbitration and has the burden of proof to support their claims; the defendant will be called to respond and to prove that they acted according to the agreement. The dispute is generally mediated, if at all, by community administrators or moderators. Eventually, the parties could settle on a new agreement or, in case of no deal, the party found guilty by the arbitrator is punished with a loss in reputation or with a ban from the community~\cite{georgoulias2021qualitative}. It is worth noting that the mere presence of an arbitration section in a market may not be enough; the mediators should intervene to take action swiftly and should be impartial, as a biased mediator could render an unfair verdict~\cite{allodi2016then}.

\subsubsection{Reputation systems}
\label{subsubsec:framework_reputation}
Reputation systems aim at creating conditions in which users with a higher reputation are perceived by other users as being more trustworthy, resulting in a greater willingness to trade~\cite{leukfeldt2015organised}. In other words, reputation systems help mitigate information asymmetry problems~\cite{dupont2017darkode, dupont2016ecology, tabarrok2015end, yip2013trust} by indicating the quality of the offered products, thus easing customers in assessing the products' properties ahead of purchase~\cite{yip2013forums, huang2018systematically}. Similarly, reputation systems can serve as a mechanism to `punish' users who misbehave~\cite{soska2015measuring}, potentially discouraging dishonest members from participating in trade. That is particularly important as it promotes predictable, `honest' behavior in the community (as opposed to capitalizing on misbehavior)~\cite{georgoulias2021qualitative, lusthaus2012trust}, and may enable a more aggressive pricing strategy from reputable members~\cite{yip2013forums}, and hence greater economic returns. Similarly to restricted access models, obtaining a privileged status in a community may be subject to the payment of a fee or the scrutiny of the community administrators~\cite{georgoulias2021qualitative, thomas2015framing}. For example, administrators may require the provision of product samples to verify their quality~\cite{hutchings2015crime, georgoulias2021qualitative, thomas2015framing} and apply different fees according to the type of product to sell.
On the other hand, the literature shows that reputation systems can also be abused by market participants~\cite{soudijn2012cybercrime, thomas2015framing}. For example, in Sybil attacks, a user creates multiple identities in a forum to artificially increase the number of positive reviews associated with their main identity. Forum mechanisms can be put in place to alleviate this risk, for example, by allowing providing positive/negative feedback on a user reputation only after trade~\cite{thomas2015framing}. 

\subsubsection{Mitigation of perverse incentives}
\label{subsubsec:framework_perverse}

In underground communities, the problems of trust are not only limited to the interaction among peers. Administrators are oftentimes in the powerful position of controlling user funds (e.g., through escrow services provided by the market) while having to deal with the cost of running a marketplace, both in terms of efforts and economic expenses~\cite{huang2018systematically}. The presence of clear and transparent revenue streams for market administrators can play a role in discouraging actions such as exit scams. Revenue may be generated, for example, from the payment of fees to earn a status for both regular users and sellers~\cite{georgoulias2021qualitative}, transaction costs or fees~\cite{georgoulias2021qualitative, lusthaus2012trust}, and by providing advertisement space on the platform~\cite{georgoulias2021qualitative, huang2018systematically}.

\subsubsection{Rule system}
\label{subsubsec:framework_rules}

Next to regulatory mechanisms to establish trust in trade, markets can provide a list of enforceable rules to guide trade activities on the market~\cite{dupont2017darkode, lusthaus2012trust} and present a hierarchical structure with members in charge of enforcing those by moderating content and taking actions against offenders~\cite{yip2013trust}. Rules can specify how contracts and agreements are enforced and who (and which users in a forum) are the authorities acting in case of non-compliance~\cite{lusthaus2012trust, schelling1980strategy}. Rules can be relatively specific, for example stating specific procedures to advertise products and engage in trade, or very loose (e.g., `scamming is not allowed').

\subsubsection{Strong authentication and anti-bot features}
\label{subsubsec:framework_authentication}

Law enforcement agencies, threat intelligence operators, and researchers closely observe the activity of underground communities, often with the support of automated crawlers to extract and aggregate data on ongoing criminal operations. It is not uncommon to find messages from community administrators warning about the presence of law enforcement and bots whose purpose is to monitor activities~\cite{vx-underground_2022}. Sometimes, law enforcement operators clearly state their presence on the market as a deterrent~\cite{abrams_2021}. Market operators employ a range of techniques to limit these undesired activities~\cite{turk2020tight, campobasso2022threat, decary2015sifting}; for example, they may employ CAPTCHAs to hinder automated monitoring and may offer their members strong authentication mechanisms limiting unwanted access to user accounts.

\subsubsection{Escrow systems}
\label{subsubsec:framework_escrow}

Numerous marketplaces employ trade protection mechanisms, such as escrow services~\cite{soska2015measuring, lusthaus2012trust, laferriere2023examining, georgoulias2021qualitative, yip2013trust}, to prevent users from falling victim to fraud. Generally, a party external to the trade (e.g., a market admin) acts as the escrow service. This party may hold payments until both parties confirm the other has fulfilled their part of the contract. Community users are generally free to use their preferred escrow services; however, some communities actively promote or provide escrow services on their platform to keep complete control in case of dispute (and potentially to earn a commission). 

\subsubsection{Interaction model}
\label{subsubsec:framework_interaction}

How users interact by writing and reading feedback or comments on products can significantly affect the availability of related information~\cite{laferriere2023examining, decary2016criminals}. 
Most forums allow unconstrained interactions in the public space of the forum, allowing buyers to ask for additional information from sellers, and to send positive or negative signals to their peers regarding specific products or users~\cite{yip2013trust}. In some systems, parts of this information are hidden and only available after payment of a fee (paid with a forum currency or with the account balance). Further, some markets encourage or provide private channels for users to interact outside of the public eye~\cite{allodi2016then, yip2013forums}.

\subsection{Framework construction}
\label{subsec:framework_costruction}

\begin{figure*}
    \centering
    \includegraphics[width=0.96\textwidth]{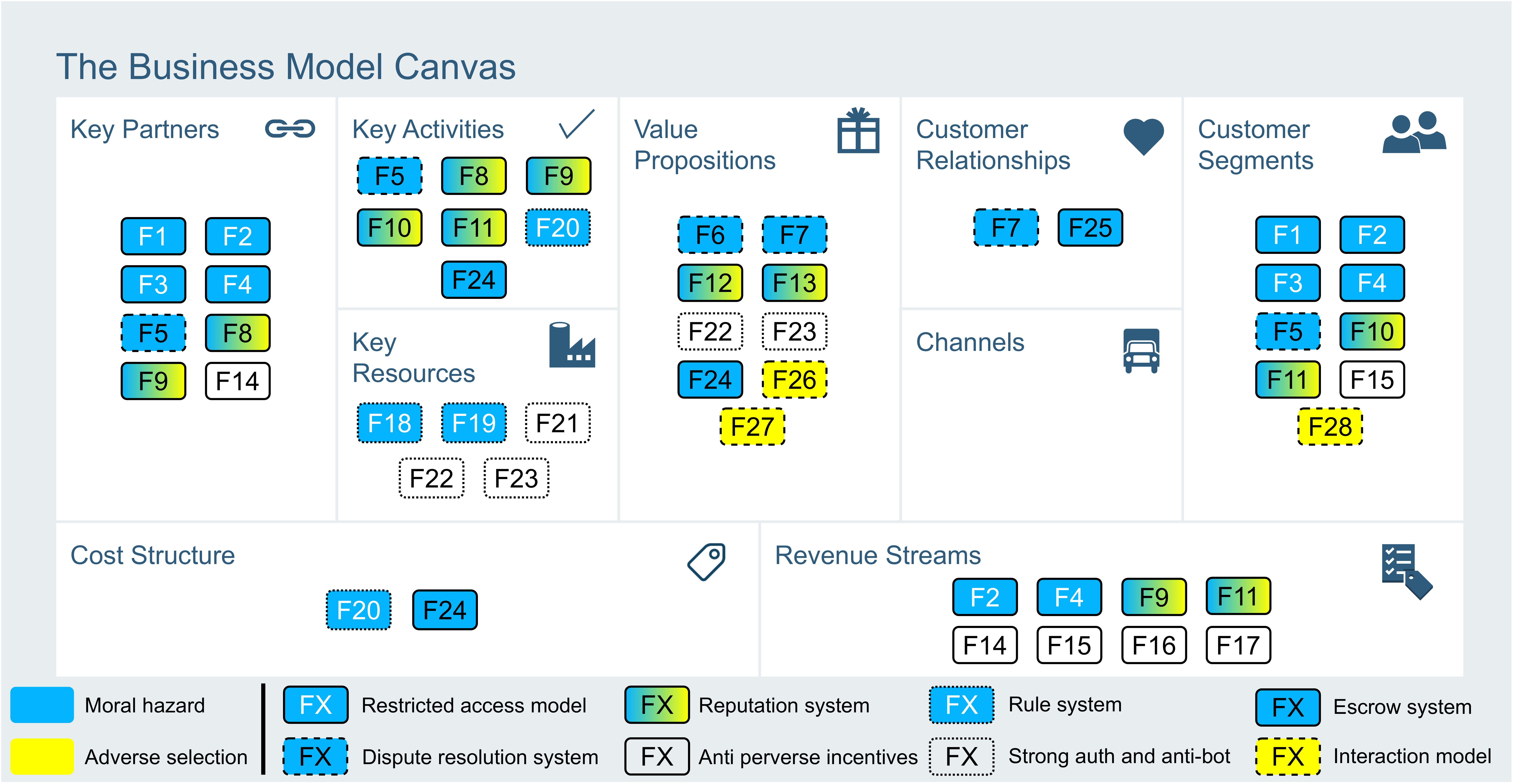}
    \caption{Business Model Canvas populated with the identified market features.}
    \begin{minipage}{\textwidth}
    \footnotesize

    \begin{multicols}{4}
    \textbf{Restricted access model}\\
    \textbf{F1.} Restricted sections - pull-in \\
    \textbf{F2.} Restricted sections - payment \\
    \textbf{F3.} Restricted registration - pull-in \\
    \textbf{F4.} Restricted registration - payment
    
    \smallskip \noindent
    \textbf{Dispute resolution system}\\
    \textbf{F5.} Scammers banned \\
    \textbf{F6.} Working dispute resolution system \\ 
    \textbf{F7.} Neutral mediator

    \smallskip \noindent
    \textbf{Reputation system}\\
    \textbf{F8.} Seller status - verification \\
    \textbf{F9.} Seller status - payment \\
    \textbf{F10.} User status - verification \\
    \textbf{F11.} User status - payment \\
    \textbf{F12.} Reputation change on trade \\
    \textbf{F13.} Reputation change by VIP

    \smallskip \noindent
    \textbf{Anti perverse incentives}\\
    \textbf{F14.} Seller status - recurrent fee \\
    \textbf{F15.} User status - recurrent fee \\
    \textbf{F16.} Escrow fee \\
    \textbf{F17.} Sponsored ads
    
    \smallskip \noindent
    \textbf{Rule system}\\
    \textbf{F18.} Clear trade rules \\
    \textbf{F19.} Moderators roles \\
    \textbf{F20.} Active moderation

    \smallskip \noindent
    \textbf{Strong authentication and anti-bot}\\
    \textbf{F21.} 2FA available \\
    \textbf{F22.} CAPTCHA on authentication \\
    \textbf{F23.} CAPTCHA on access

    \smallskip \noindent
    \textbf{Escrow system}\\
    \textbf{F24.} Escrow available \\
    \textbf{F25.} Escrow recommended

    \smallskip \noindent
    \textbf{Interaction model}\\
    \textbf{F26.} Public interaction \\
    \textbf{F27.} Private interaction \\
    \textbf{F28.} Pay to show content
    \end{multicols}
    \end{minipage}
    \label{fig:bmc}
\end{figure*}

An enumeration of the identified features is provided in Figure~\ref{fig:bmc}, showing their mapping to both the BMC and to the foundational dimensions identified in Section~\ref{sec:background} (\textit{Market participants:} moral hazard, adverse selection; \textit{Market administration:} cost structure and risks. 
As a visual guideline, the left-hand side of the BMC (Key Partners, Key Resources, and Key Activities) identifies the `needs' of the business model examined (that is, the suppliers, the physical or intellectual resources, and the activities that support the business). 
The right-hand side of the canvas identifies business aspects for Customer Relationships, Channels, and Customer Segments, which identify the target of the business examined, how to reach new customers, and how to maintain relationships with existing ones. 
In the middle, the Value Proposition represents the added value of the enterprise, in the form of a product or service created and delivered to customers. 
Finally, at the bottom of the canvas, the Cost Structure and Revenue Streams provide a breakdown of the cash flows of the business. 
For the sake of brevity, and because most of the identified features are self-explanatory, definitions for each feature are reported in the Appendix.
Following, a brief explanation of the mapping rationale of the market mechanisms to the BMC. 
To build the association, we follow the guidelines reported in Section~\ref{subsec:methodology_framework_derivation}.

\smallskip
\noindent
\textbf{Key Partners, Key Activities, Key Resources.} This area mostly covers the mechanisms employed by markets to select their Key Partners (sellers) by employing a \textit{restricted access model}, punish them in case of misbehavior via a dispute resolution system, in accordance to trade rules (\textit{rule system}) and give its participants a \textit{reputation system} to identify honest sellers selling quality products. To further limit moral hazard, a market could provide an \textit{escrow system}. To improve their Value Proposition, markets could include in their Key Resources means to mitigate the risks of account hijacking attacks and to provide more privacy to their participants, such as \textit{strong authentication} and the implementation of \textit{anti-bot features}. It is worth noting that guaranteeing market administrations a constant influx of money via the payment of recurrent fees from market participants is beneficial to mitigate \textit{perverse incentives}.

\smallskip
\noindent
\textbf{Customer Relationships, Customer Segments, Channels.}
Features related to buyers and the establishment of trust relationships are mapped to the right side of the BMC. Similarly to the previous paragraph, markets can identify their Customer Segments by screening and selecting market participants with the use of strictly \textit{restricted access models}. Honest participants could benefit from a higher \textit{reputation}; markets could decide whether this status should be earned by generic \textit{interactions} with other members or as a result of trade thus mitigating the problems related to adverse selection, or with the payment of a (\textit{recurrent}) fee, thus supporting the market over time and mitigating \textit{perverse incentives}. Markets commit to their customers and the general health of the community by offering a \textit{dispute resolution system} that impartially punishes dishonest members and provides guidelines and recommendations for the use of \textit{escrow systems}. 

\smallskip
\noindent
\textbf{Value Proposition.}
Depending on how a market screened its participants, the resources allocated and the activities performed to support the business, and the relationship it has with its customers, a different \textit{Value Proposition} may emerge. A market could offer its participants a working and impartial \textit{dispute resolution system}, together with an \textit{escrow system} that guarantees that the honest party obtains their money in case of disagreement. Also, a market could implement a transparent \textit{reputation system} to promote quality \textit{interactions} to create signals among peers and could commit to keeping unwanted visitors out of it with the implementation of \textit{anti-bot mechanisms}. 

\smallskip
\noindent
\textbf{Cost Structure, Revenue Streams.}
Features related to revenue streams and administration risks are grouped under Cost Structure and Revenue Streams. Intuitively, a market's revenue depends on the presence of (recurrent) fees for both sellers and buyers. Some fees could be imposed during the screening of new market participants (\textit{restricted access model}), while others in exchange for privileged statuses (\textit{reputation system}). Finally, the implementation of smaller, periodic fees for services and advertisement provides a market's administration with a constant influx of money, creating revenue and thus offering incentives to keep the market functional and attractive for an ever-growing pool of members, mitigating \textit{perverse incentives}.


\section{Results} 
\label{sec:results}

\subsection{`Hits' within our market selection}
\label{subsec:results_aliases}

\begin{figure}
    \centering
    \includegraphics[width=0.92\columnwidth]{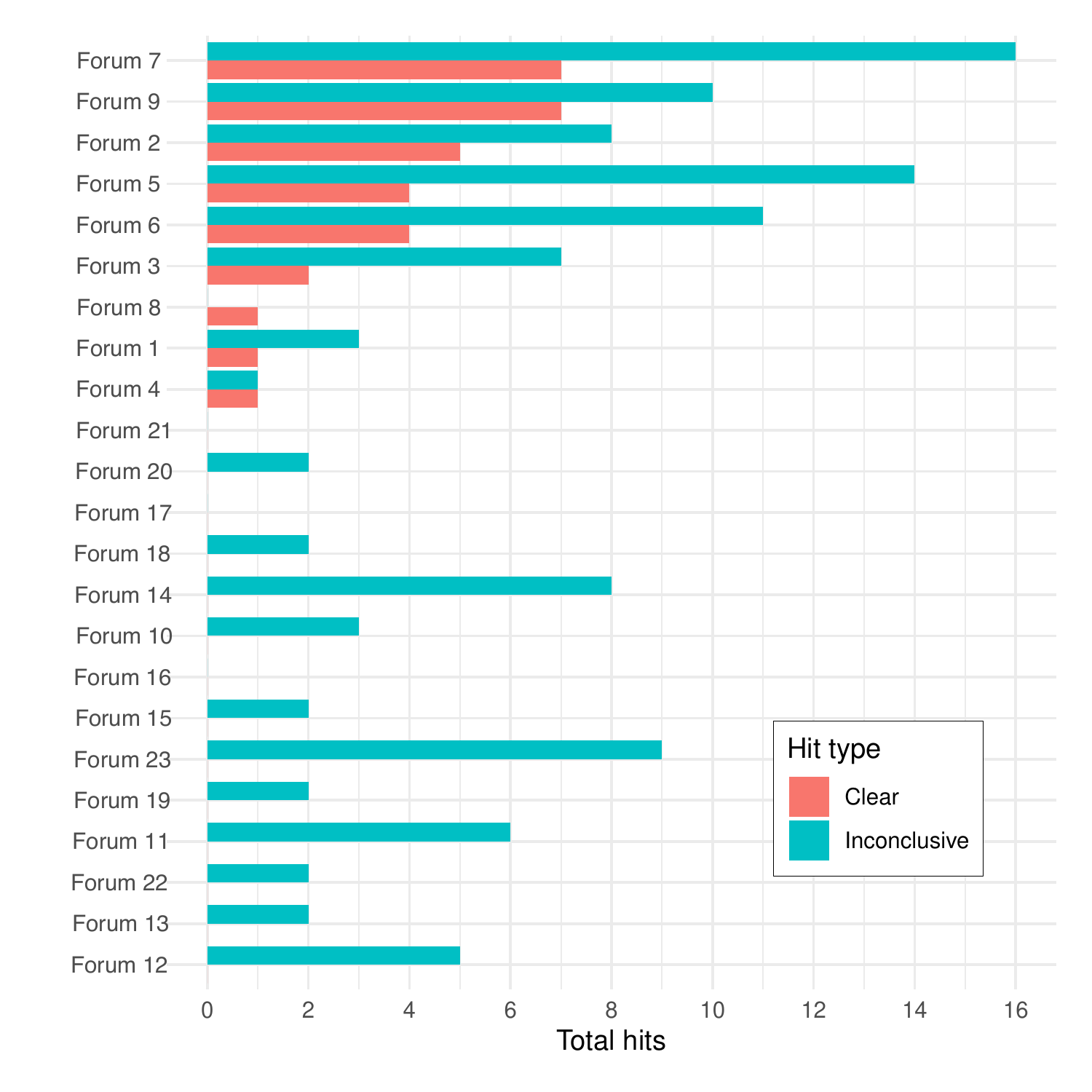}
    \caption{The number of clear and inconclusive hits over criminal communities}
    \label{fig:clearhit}
\end{figure}

From a total of 836 aliases associated with convicted cybercriminals, 380 had at least one account on one of the analyzed forums with an identical name. Thus, 456 aliases are automatically labeled as no hits. The remaining 380 aliases allowed us to identify, across all forums, 32 instances that resulted in clear hits (25 distinct aliases), 113 instances of accounts that resulted in inconclusive hits (accounting for 88 distinct aliases), while the remaining 225 aliases resulted in accounts with unrelated activity or no activity whatsoever (no hits). Figure~\ref{fig:clearhit} 
reports an overview of the distribution of the hits across the examined forums. Among the clear hits, three aliases were identified as belonging to the same person in three or more forums. \omerta, \exploit\ and \crdclub\ account for $\approx60\%$ of clear hits we could identify. An interesting aspect shared across the marketplaces with the highest number of clear hits is longevity: \omerta, \exploit, \crdclub\ and \verified\ are longstanding underground communities that have been operating for 9 years or more. This could possibly indicate that their success as criminal venues comes from having convincingly addressed trust issues over multiple years of experience. Observing the top four markets for inconclusive hits (\omerta, \verified, \hackforums, and \exploit), accounting for  $\approx45\%$ of them, we note that they also feature a rather high number of clear hits ($\approx70\%$), possibly indicating that some content from those members labeled as inconclusive hits could have been deleted. In fact, interestingly, the forums \crdclub, \verified\ and \omerta\ seem to have performed a shadow-banning of some of their members (respectively 9, 9, and 16 aliases); when searched for them, we were shown a different error page than the one shown when the username or associated content could not be found. For all three forums, we could identify the URL that would have displayed the page associated with the shadow-banned alias; we accessed it and labeled the findings accordingly. It is worth noting that the three markets shadow-banning aliases are among those that scored the highest number of hits. That suggests that their administrators are aware of their exposure and take actions to mitigate collateral damage following law enforcement operations. Similarly to shadow-banning, \crdclub, \enclave, \omerta, \verified\ and \crackingpro\ seem to have purged content from the searched accounts in 7, 6, 3, 1, and 1 cases, respectively. In fact, for the investigated forums, simply banning an account does not remove the content posted by that author. We noted the absence of content in some profiles from the mismatch in the count of posts associated with each account and the posts we could find, and we verified that the discrepancy could not be caused by the removal of content by the authors themselves, as it is not allowed in any of the platforms. In this case, however, in the absence of content, these aliases were labeled as no hits. 

\subsection{Market features}
\label{subsec:results_labelling}

\begin{table*}
\centering
\scalebox{0.54}{
\begin{tabular}{p{0.156\textwidth}|p{0.008\textwidth}p{0.008\textwidth}p{0.008\textwidth}p{0.008\textwidth}p{0.008\textwidth}p{0.008\textwidth}p{0.008\textwidth}p{0.015\textwidth}|p{0.015\textwidth}p{0.016\textwidth}p{0.012\textwidth}p{0.012\textwidth}p{0.015\textwidth}|p{0.008\textwidth}p{0.008\textwidth}p{0.008\textwidth}p{0.015\textwidth}p{0.015\textwidth}p{0.015\textwidth}p{0.016\textwidth}|p{0.008\textwidth}p{0.008\textwidth}p{0.012\textwidth}p{0.012\textwidth}p{0.012\textwidth}p{0.012\textwidth}p{0.016\textwidth}p{0.016\textwidth}p{0.016\textwidth}|p{0.023\textwidth}p{0.023\textwidth}|p{0.008\textwidth}p{0.008\textwidth}p{0.008\textwidth}p{0.008\textwidth}p{0.008\textwidth}p{0.012\textwidth}p{0.015\textwidth}p{0.016\textwidth}p{0.016\textwidth}|p{0.008\textwidth}p{0.008\textwidth}p{0.008\textwidth}p{0.012\textwidth}p{0.015\textwidth}p{0.015\textwidth}p{0.015\textwidth}p{0.015\textwidth}|p{0.022\textwidth}p{0.022\textwidth}}
\toprule
& \multicolumn{8}{c|}{\textbf{Key Partners}} & \multicolumn{5}{c|}{\textbf{Key Resources}} & \multicolumn{7}{c|}{\textbf{Key Activities}} & \multicolumn{9}{c|}{\textbf{Value Proposition}} & \multicolumn{2}{c|}{\textbf{Cst. Rel.}} & \multicolumn{9}{c|}{\textbf{Customer Segments}} & \multicolumn{8}{c|}{\textbf{Revenue Streams}} & \multicolumn{2}{c}{\textbf{Cost Str.}} \\
\midrule
& F1 & F2 & F3 & F4 & F5 & F8 & F9 & F14 & F18 & \cellcolor[HTML]{C0C0C0}F19 & F21 & F22 & F23 & \cellcolor[HTML]{FFFF99}F5 & F8 & F9 & F10 & F11 & F20 & F24 & F6 & \textbf{F7} & F12 & F13 & F22 & F23 & F24 & \cellcolor[HTML]{C0C0C0}F26 & F27 & F7 & F25 & F1 & F2 & F3 & F4 & F5 & F10 & F11 & \cellcolor[HTML]{FFFF99}F15 & F28 & F2 & F4 & F9 & F11 & F14 & \cellcolor[HTML]{FFFF99}F15 & \cellcolor[HTML]{C0C0C0}F16 & F17 & F20 & F24 \\ \midrule
\raidforums& \fno & \fyes & \fno & \fno & \fyes & \fno & \fno & \fno  & \fno &-& \funkn & \fyes & \fyes  & \fyes & \fno & \fno & \fno & \fyes & \fyes  & \fyes & \fyes & \fyes & \fyes & \fno & \fyes & \fyes & \fyes & \fyes & \fno & \fyes & \fyes    & \fno & \fyes & \fno & \fno & \fyes & \fno & \fyes & \fno & \fyes  & \fyes & \fno & \fno & \fyes & \fno & \fno & - & \fyes & \fyes & \fyes \\
\crdclub& \fyes & \fyes & \fno & \fno & \fyes & \fyes & \fyes & \fyes  & \fno &-& \fno & \fno & \fno  & \fyes & \fyes & \fyes & \fyes & \fno & \fyes  & \fyes & \fyes & \fyes & \fno & \fyes & \fno & \fno & \fyes & \fyes & \fyes & \fyes & \fno    & \fyes & \fyes & \fno & \fno & \fyes & \fyes & \fno & \fno & \fno  & \fyes & \fno & \fyes & \fno & \fyes & \fno & - & \fyes & \fyes & \fyes \\
\xss& \fyes & \fyes & \fno & \fno & \fyes & \fno & \fyes & \fyes  & \fyes &-& \fyes & \fno & \fno  & \fyes & \fno & \fyes & \fno & \fno & \fyes  & \fyes & \fyes & \fyes & \fno & \fno & \fno & \fno & \fyes & \fyes & \fyes & \fyes & \fyes    & \fyes & \fyes & \fno & \fno & \fyes & \fno & \fno & \fno & \fno  & \fyes & \fno & \fyes & \fno & \fyes & \fno & - & \fyes & \fyes & \fyes \\
\fuckav& \fno & \fno & \fno & \fno & \fyes & \fno & \funkn & \funkn  & \funkn &-& \funkn & \fno & \fno  & \fyes & \fno & \funkn & \fno & \funkn & \fyes  & \funkn & \fyes & \fyes & \fno & \fno & \fno & \fno & \funkn & \fyes & \fyes & \fyes & \fno    & \fno & \fno & \fno & \fno & \fyes & \fno & \funkn & \funkn & \funkn  & \fno & \fno & \funkn & \funkn & \funkn & \funkn & - & \fyes & \fyes & \funkn \\
\verified& \fyes & \fyes & \fno & \fyes & \fyes & \fyes & \fyes & \fno  & \fyes &-& \fyes & \fno & \fyes  & \fyes & \fyes & \fyes & \fno & \fyes & \fyes  & \fyes & \fyes & \fyes & \fyes & \fno & \fno & \fyes & \fyes & \fyes & \fyes & \fyes & \fyes    & \fyes & \fyes & \fno & \fyes & \fyes & \fno & \fyes & \fyes & \fno  & \fyes & \fyes & \fyes & \fyes & \fno & \fyes & - & \fyes & \fyes & \fyes \\
\hackforums& \fno & \fno & \fno & \fno & \fyes & \fyes & \fyes & \fno  & \fyes &-& \fyes & \fno & \fno  & \fyes & \fyes & \fyes & \fno & \fyes & \fyes  & \fno & \fyes & \fno & \fyes & \fno & \fno & \fno & \fno & \fyes & \fyes & \fno & \fno    & \fno & \fno & \fno & \fno & \fyes & \fno & \fyes & \fyes & \fyes  & \fno & \fno & \fyes & \fyes & \fno & \fyes & - & \fyes & \fyes & \fno \\
\omerta& \funkn & \funkn & \fno & \fyes & \fno & \fno & \fyes & \fno  & \fno &-& \fno & \fno & \fno  & \fno & \fno & \fyes & \fno & \funkn & \fno  & \fyes & \fno & \fno & \fno & \fno & \fno & \fno & \fyes & \fyes & \fyes & \fno & \fyes    & \funkn & \funkn & \fno & \fyes & \fno & \fno & \funkn & \funkn & \fno  & \funkn & \fyes & \fyes & \funkn & \fno & \funkn & - & \fyes & \fno & \fyes \\
\enclave& \fno & \fno & \fno & \fno & \fno & \fno & \fyes & \funkn  & \fno &-& \fno & \fno & \fno  & \fno & \fno & \fyes & \fno & \fno & \fno  & \fno & \fno & \fno & \fno & \fno & \fno & \fno & \fno & \fyes & \fyes & \fno & \fno    & \fno & \fno & \fno & \fno & \fno & \fno & \fno & \fno & \fno  & \fno & \fno & \fyes & \fno & \funkn & \fno & - & \fyes & \fno & \fno \\
\exploit& \fyes & \fno & \fyes & \fyes & \fyes & \fno & \funkn & \funkn  & \fyes&-& \fyes & \fno & \fno  & \fyes & \fno & \funkn & \fno & \fno & \fyes  & \fyes & \fyes & \fyes & \fno & \fno & \fno & \fno & \fyes & \fyes & \fyes & \fyes & \fyes    & \fyes & \fno & \fyes & \fyes & \fyes & \fno & \fno & \fno & \fno  & \fno & \fyes & \funkn & \fno & \funkn & \fno & - & \fyes & \fyes & \fyes \\
\midrule
\deeptor& \fno & \fyes & \fno & \fno & \fno & \funkn & \funkn & \funkn  & \fno &-& \fyes & \fno & \fno  & \fno & \funkn & \funkn & \fno & \funkn & \fno  & \fyes & \fno & \fno & \fno & \fno & \fno & \fno & \fyes & \fyes & \fno & \fno & \fno    & \fno & \fyes & \fno & \fno & \fno & \fno & \funkn & \funkn & \fno  & \fyes & \fno & \funkn & \funkn & \funkn & \funkn & - & \fyes & \fno & \fyes \\
\blackhatworld& \fyes & \fyes & \fno & \fno & \fyes & \fno & \fyes & \fyes  & \fyes &-& \fyes & \fno & \fyes  & \fyes & \fno & \fyes & \fno & \fyes & \fyes & \fno   & \fyes & \fyes & \fyes & \fno & \fno & \fyes & \fno & \fyes & \fyes & \fyes & \fno    & \fyes & \fyes & \fno & \fno & \fyes & \fno & \fyes & \fyes & \fno  & \fyes & \fno & \fyes & \fyes & \fyes & \fyes & - & \fyes & \fyes & \fno \\
\torum& \fno & \fno & \fno & \fno & \fno & \fno & \fno & \fno  & \funkn &-& \funkn & \funkn & \funkn  & \fno & \fno & \fno & \fno & \fno & \funkn  & \fno & \fno & \fno & \fno & \fno & \funkn & \funkn & \fno & \fyes & \fyes & \fno & \fno    & \fno & \fno & \fno & \fno & \fno & \fno & \fno & \fno & \fno  & \fno & \fno & \fno & \fno & \fno & \fno & - & \fno & \funkn & \fno \\
\antichat& \fyes & \fno & \fno & \fno & \fyes & \fno & \fno & \fno  & \fyes&-& \fyes & \fno & \fyes  & \fyes & \fno & \fno & \fyes & \fno & \fyes  & \fyes & \fno & \fno & \fyes & \fyes & \fno & \fyes & \fyes & \fyes & \fyes & \fno & \fyes    & \fyes & \fno & \fno & \fno & \fyes & \fyes & \fno & \fno & \fno  & \fno & \fno & \fno & \fno & \fno & \fno & - & \fyes & \fyes & \fyes \\
\nulled& \fno & \fyes & \fno & \fno & \fyes & \fno & \funkn & \funkn  & \fno &-& \fyes & \fyes & \fyes  & \fyes & \fno & \funkn & \fno & \fyes & \fyes  & \fyes & \fyes & \fno & \fyes & \fyes & \fyes & \fyes & \fyes & \fyes & \fyes & \fno & \fyes    & \fno & \fyes & \fno & \fno & \fyes & \fno & \fyes & \fyes & \fno  & \fyes & \fno & \funkn & \fyes & \funkn & \fyes & - & \fyes & \fyes & \fyes \\
\crackingpro& \fno & \fyes & \fno & \fno & \fno & \fno & \fyes & \fyes  & \fno &-& \fno & \fno & \fno  & \fno & \fno & \fyes & \fno & \fyes & \fno  & \fno & \fno & \fno & \fno & \fno & \fno & \fno & \fno & \fyes & \fyes & \fno & \fno    & \fno & \fyes & \fno & \fno & \fno & \fno & \fyes & \fyes & \fno  & \fyes & \fno & \fyes & \fyes & \fyes & \fyes & - & \fyes & \fno & \fno \\
\cryptbb& \fno & \fno & \fyes & \fno & \fno & \fyes & \fno & \fno  & \fno &-& \fno & \fyes & \fno  & \fno & \fyes & \fno & \fno & \fno & \fno  & \fyes & \fno & \fno & \fno & \fno & \fyes & \fno & \fyes & \fyes & \fyes & \fno & \fyes    & \fno & \fno & \fyes & \fno & \fno & \fno & \fno & \fno & \fno  & \fno & \fno & \fno & \fno & \fno & \fno & - & \fyes & \fno & \fyes \\
\altenen& \fno & \fyes & \fno & \fno & \fyes & \fno & \fyes & \fno  & \fno &-& \fyes & \fyes & \fno  & \fyes & \fno & \fyes & \fno & \fyes & \fyes  & \fyes & \fyes & \fno & \fyes & \fno & \fyes & \fno & \fyes & \fyes & \fyes & \fno & \fyes    & \fno & \fyes & \fno & \fno & \fyes & \fno & \fyes & \fyes & \fno  & \fyes & \fno & \fyes & \fyes & \fno & \fyes & - & \fyes & \fyes & \fyes \\
\rutor& \fno & \fyes & \fno & \fno & \fyes & \fno & \fyes & \fno  & \fyes&-& \fyes & \fno & \fyes  & \fyes & \fno & \fyes & \fno & \fyes & \fyes  & \fyes & \fyes & \fyes & \fno & \fno & \fno & \fyes & \fyes & \fyes & \fno & \fyes & \fyes    & \fno & \fyes & \fno & \fno & \fyes & \fno & \fyes & \fno & \fno  & \fyes & \fno & \fyes & \fyes & \fno & \fno & - & \fyes & \fyes & \fyes \\
\cebulka& \fyes & \fno & \fno & \fno & \fyes & \fyes & \fyes & \fno  & \fno &-& \fyes & \fno & \fyes  & \fyes & \fyes & \fyes & \fyes & \fyes & \fno  & \fyes & \fyes & \fno & \fyes & \fyes & \fno & \fyes & \fyes & \fyes & \fyes & \fno & \fyes    & \fyes & \fno & \fno & \fno & \fyes & \fyes & \fyes & \fno & \fno  & \fno & \fno & \fyes & \fyes & \fno & \fno & - & \fno & \fno & \fyes \\
\darknetcity& \fno & \fno & \fno & \fyes & \fyes & \funkn & \funkn & \funkn  & \funkn  &-& \funkn & \funkn & \funkn  & \fyes & \funkn & \funkn & \funkn & \fno & \funkn  & \fyes & \fyes & \fno & \funkn & \funkn & \funkn & \funkn & \fyes & \fyes & \fyes & \fno & \fyes    & \fno & \fno & \fno & \fyes & \fyes & \funkn & \fno & \funkn & \funkn  & \fno & \fyes & \funkn & \fno & \funkn & \funkn & - & \funkn & \funkn & \fyes \\
\reverse& \fno & \fno & \fno & \fno & \fno & \fno & \fno & \fno  & \fno &-& \fyes & \fno & \fno  & \fno & \fno & \fno & \fno & \fno & \fno  & \fno & \fno & \fno & \fno & \fno & \fno & \fno & \fno & \fyes & \fyes & \fno & \fno    & \fno & \fno & \fno & \fno & \fno & \fno & \fno & \fno & \fno  & \fno & \fno & \fno & \fno & \fno & \fno & - & \fno & \fno & \fno \\
\bhf& \fno & \fno & \fno & \fno & \fyes & \fno & \fyes & \fyes  & \fyes&-& \fyes & \fyes & \fno  & \fyes & \fno & \fyes & \fno & \fyes & \fyes  & \fyes & \fyes & \fno & \fno & \fno & \fyes & \fno & \fyes & \fyes & \fyes & \fno & \fno    & \fno & \fno & \fno & \fno & \fyes & \fno & \fyes & \fyes & \fno  & \fno & \fno & \fyes & \fyes & \fyes & \fyes & - & \fyes & \fyes & \fyes \\
\cracked& \fno & \fyes & \fno & \fno & \fyes & \fno & \fno & \fno  & \fno&-& \fyes & \fyes & \fyes  & \fyes & \fno & \fno & \fno & \fyes & \fyes  & \fyes & \fyes & \fno & \fyes & \fyes & \fyes & \fyes & \fyes & \fyes & \fyes & \fno & \fyes    & \fno & \fyes & \fno & \fno & \fyes & \fno & \fyes & \fyes & \fno  & \fyes & \fno & \fno & \fyes & \fno & \fyes & - & \fyes & \fyes & \fyes \\
\midrule
Successful markets & \multicolumn{8}{c|}{\textbf{46.15\%}} & \multicolumn{5}{c|}{33.33\%} & \multicolumn{7}{c|}{53.06\%} & \multicolumn{9}{c|}{47.89\%} & \multicolumn{2}{c|}{\textbf{61.11\%}} & \multicolumn{9}{c|}{37.31\%} & \multicolumn{8}{c|}{58.70\%} & \multicolumn{2}{c}{76.47\%} \\
Unsuccessful markets & \multicolumn{8}{c|}{30.77\%} & \multicolumn{5}{c|}{\textbf{52.08\%}} & \multicolumn{7}{c|}{46.67\%} & \multicolumn{9}{c|}{50.00\%} & \multicolumn{2}{c|}{\textbf{35.71\%}} & \multicolumn{9}{c|}{23.16\%} & \multicolumn{8}{c|}{46.05\%} & \multicolumn{2}{c}{65.38\%} \\
\bottomrule
\end{tabular}
}
\vspace{1mm}
\caption{Overview of the framework applied to the selected markets.}
    \vspace{-3mm}
    \begin{minipage}{0.9\textwidth}
        \footnotesize The upper part of the table contains the communities that have at least one clear hit, while the lower part contains the communities that have no clear hit. We use the following encoding: \textbf{-} if the feature does not exist on the respective community, \checkmark if the community exhibits it, and \textbf{?} if we were unable to find whether the feature exists or not on the respective community. We highlight the features excluded from the computed fraction of available features per group: in gray, the non-binary and constant dimensions; in yellow, the dimensions causing multi-collinearity. 
    \end{minipage}
    \begin{minipage}{\textwidth}
    \footnotesize
    \begin{multicols}{4}
    \textbf{Key Partners}\\
    \textbf{F1.} Restricted sections - pull-in \\
    \textbf{F2.} Restricted sections - payment \\
    \textbf{F3.} Restricted registration - pull-in \\
    \textbf{F4.} Restricted registration - payment \\
    \textbf{F5.} Scammers banned \\    
    \textbf{F8.} Seller status - verification \\
    \textbf{F9.} Seller status - payment \\
    \textbf{F14.} Seller status - recurrent fee
    
    \smallskip \noindent
    \textbf{Key Resources}\\
    \textbf{F18.} Clear trade rules \\
    \textbf{F19.} Moderators roles \\
    \textbf{F21.} 2FA available \\
    \textbf{F22.} CAPTCHA on authentication \\
    \textbf{F23.} CAPTCHA on access

    \smallskip \noindent
    \textbf{Key Activities}\\
    \textbf{F5.} Scammers banned \\
    \textbf{F8.} Seller status - verification \\
    \textbf{F9.} Seller status - payment \\
    \textbf{F10.} User status - verification \\
    \textbf{F11.} User status - payment \\
    \textbf{F20.} Active moderation \\
    \textbf{F24.} Escrow available

    \smallskip \noindent
    \textbf{Value Proposition}\\
    \textbf{F6.} Working dispute resolution system \\ 
    \textbf{F7.} Neutral mediator \\
    \textbf{F12.} Reputation change on trade \\
    \textbf{F13.} Reputation change by VIP \\
    \textbf{F22.} CAPTCHA on authentication \\
    \textbf{F23.} CAPTCHA on access \\
    \textbf{F24.} Escrow available \\
    \textbf{F26.} Public interaction \\
    \textbf{F27.} Private interaction

    \smallskip \noindent
    \textbf{Customer Relationships}\\
    \textbf{F7.} Neutral mediator \\
    \textbf{F25.} Escrow recommended

    \smallskip \noindent
    \textbf{Customer Segments}\\
    \textbf{F1.} Restricted sections - pull-in \\
    \textbf{F2.} Restricted sections - payment \\
    \textbf{F3.} Restricted registration - pull-in \\
    \textbf{F4.} Restricted registration - payment \\
    \textbf{F5.} Scammers banned \\ 
    \textbf{F10.} User status - verification \\
    \textbf{F11.} User status - payment \\
    \textbf{F15.} User status - recurrent fee \\
    \textbf{F28.} Pay to show content

    \smallskip \noindent
    \textbf{Revenue Streams}\\
    \textbf{F2.} Restricted sections - payment \\
    \textbf{F4.} Restricted registration - payment \\
    \textbf{F9.} Seller status - payment \\
    \textbf{F11.} User status - payment \\
    \textbf{F14.} Seller status - recurrent fee \\
    \textbf{F15.} User status - recurrent fee \\
    \textbf{F16.} Escrow fee \\
    \textbf{F17.} Sponsored ads

    \smallskip \noindent
    \textbf{Cost Structure}\\
    \textbf{F20.} Active moderation \\
    \textbf{F24.} Escrow available
    \end{multicols}
    \end{minipage}
\label{tab:features}
\end{table*}


We proceed to label the selected markets according to the framework. In Table~\ref{tab:features}, we report an overview of all communities with their corresponding features. The upper part of the table reports successful markets. The bottom of the table reports the aggregated fractions of features available per group (as identified from the BMC), calculated both for successful and unsuccessful markets. For the purposes of the analysis and to discuss their relevance, we do not include in the said fraction non-binary features (F16 - escrow fee, F19 - moderator roles) and variables with constant values (F26 - public interaction). Furthermore, we compute the correlation matrices to identify potential multi-collinearity across features within each group and to remove the problematic ones.   
Among the 23 selected markets, 4 (\raidforums, \fuckav, \torum\ and \darknetcity) are no longer reachable. For \darknetcity, we could rely on partial labeling of the market we performed at a prior version of the framework, and for which we already investigated the available aliases; for \torum, \fuckav\ and \hackforums\ we had access to historical data provided from an industrial partner, Web-IQ, which allowed us to perform partial labeling and investigate on the presence of criminals on the market; in addition, for \fuckav\ and \hackforums, we relied on \verb|archive.org| to explore some portions of the market (e.g., community rules) that were not collected from Web-IQ. For the remaining 19 markets, we could confidently verify the presence of the investigated features for the majority of cases. Nonetheless, sometimes, even when having access to said markets, it was not possible to clearly decide on the presence of some features. 
For example, \deeptor\ features sections where `Trusted \& Verified Sellers' are allowed to post. However, we could not identify any clear mention of how to become a trusted seller in the market.

From the framework, we identify some substantial differences across groups between successful and unsuccessful markets. A set of one-tailed Fisher tests for the alternative hypothesis that successful markets have a greater count of features within a group than unsuccessful markets finds at least marginally statistically significant differences for the groups Key Partners ($p-value=0.032$), Key Resources ($p=0.074$), and Customer Relationships ($p=0.083$). In particular, the presence of features in Key Partners and Customer Relationships is more frequent for successful markets while, perhaps surprisingly, they are less frequent for Key Resources. The features collected in Key Partners indicate a greater tendency for successful markets to be more segregated, screening or vetting their community base, creating tiers of access to portions of their markets than unsuccessful markets. The importance of Customer Relationships is mostly attributable to F7: neutral mediator (which turns out to be the only statistically significant variable when considered alone; $p=0.017$), indicating that criminals may value impartiality in the administration of markets\footnote{We note that, for \rutor, we conservatively assign the \checkmark for F7 as the size of the forum made unfeasible the verification of administrators not being involved in trade activities.}. A closer look at the considered variables in Key Resources shows that CAPTCHAs are more frequent for unsuccessful markets and that they feature more often the possibility to enable 2FA, perhaps indicating a greater risk perceived from those markets to be victims of DDoS attacks and other abuses (as it is the case of markets promoting booter services, suffering attacks from their competition~\cite{santanna2015booters, karami2013understanding}).

\subsection{Market similarity}
\label{subsec:results_similarity}

\begin{figure}
    \centering
    \includegraphics[width=0.94\columnwidth]{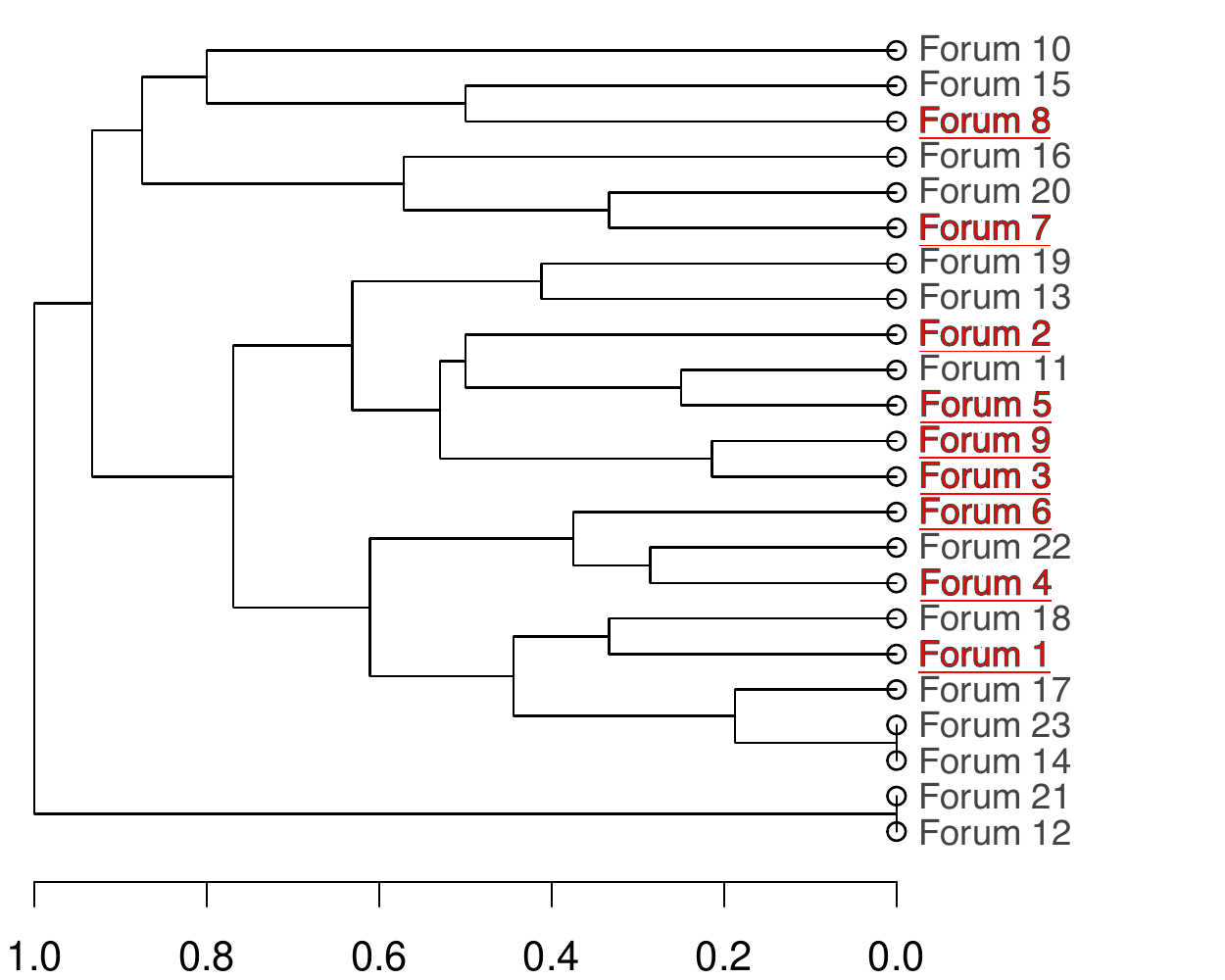}
    \caption{Hierarchical clustering of all communities based on their features. `Successful' communities are reported in red, underlined.}
    \label{fig:dendro}
\end{figure}

After analyzing the features at a group level and individually, we further investigate the similarities among the communities. We perform a hierarchical clustering of the markets with complete linkage dissimilarity and compute the Jaccard distance on their feature set.
The resulting dendrogram is depicted in Figure~\ref{fig:dendro}.
The level at which forums are connected in the dendrogram indicates how similar the analyzed communities are when compared through their labeled features (i.e., the lower the level, the higher the similarity). Communities featuring at least a clear hit (i.e., `successful') are reported in red. Features set to unknown for a given market (represented with \textbf{?}) are ignored for comparison. 

We first notice that most successful markets are relatively close in the dendrogram, with two clusters in the lower part of the figure linking at a dissimilarity level of 0.6, except for \omerta\ and \enclave. Interestingly, according to some discussion in affiliated communities, we learn that \omerta\ (a still-active, large marketplace) experienced its pinnacle of fame in the carding scene during the mid/late-2010s and now is in a descending phase. We could speculate that the classification of \omerta\ as a successful market and its current descending phase could indicate that its characteristics may have been comparatively attractive for its time. However, the lack of evolution towards other mechanisms employed by competitors moved customers and sellers to other attractive alternatives, leading us to consider it more as an outlier. \enclave\ features only one hit and seems a low-active market with little activity from the administrators. We contacted the administrators to obtain information on the benefits we would obtain by paying the `activation fee'. After more than two months, we did not receive any answer yet; when we asked another user who activated their account, they confirmed our suspicions that the payment does not grant access to any additional market section. That suggests that this could be considered more of an outlier, rather than a successful market. These considerations may justify the reason why these two markets seem less correlated to the other sub-tree.

Among the successful markets, \exploit\ and \xss\ are the most similar; the two markets differ only in terms of access mechanisms: \exploit\ vets registrations manually or via the payment of a fee, while \xss\ employs those mechanisms to limit access to a portion of the market. These two markets link at around 0.5 with \crdclub\ and \verified.
Comparatively, \crdclub\ appears to be leaner on the regulatory side, with no explicit rules or recommendations on trade, the lack of a trade-based reputation system, although sporadically offers privileged statuses to vetted members. Both \crdclub\ and \verified\ look more concerned than \exploit\ and \xss\ in vetting sellers; \verified\ offers a score associated with trade activity, and is more concerned about on the revenue streams overall. 

On the other hand, unsuccessful markets appear more spread out over the dendrogram and only connect at higher levels closer to 1. To further corroborate this observation, we compare the distances between all successful and unsuccessful markets. A Wilcoxon rank-sum test indicates that successful markets are on average more similar to each other than unsuccessful markets are ($p=0.016$). That suggests that the feature configuration of successful markets has less variability (i.e., is more stable) than it has for unsuccessful markets in our set. Similarly, we find strong evidence that successful markets are more similar to each other than they are to unsuccessful markets ($p<0.0001$); in other words, there seems to be a combination of features over which successful markets are characterized, but unsuccessful markets are not.

\subsection{Qualitative observations on market features}
\label{subsec:results_qualitative}

We also observe some differences in the features between the two types of markets. 
From the framework, it emerges that the fraction of markets implementing stricter access control policies seems higher for successful markets (F1\textsubscript{s}: 50\%, F4\textsubscript{s}: 33.33\%) than for unsuccessful ones (F1\textsubscript{u}: 21.43\%, F4\textsubscript{u}: 7.14\%). 
An interesting relation emerges between the possibility to obtain privileged statuses for sellers and regular users: these mechanisms are in place for sellers more often in successful markets (F8\textsubscript{s}: 33.33\% F9\textsubscript{s}: 85.71\%) than in unsuccessful ones (F8\textsubscript{u}: 16.67\%, F9\textsubscript{u}: 54.55\%), while for users it is true the opposite (F10\textsubscript{s}: 11.11\%, F11\textsubscript{s}: 42.86\% vs F10\textsubscript{u}: 15.38\%, F11\textsubscript{u}: 61.54\%). Another notable aspect is that successful markets more often feature working dispute resolution systems (F6\textsubscript{s}: 77.78\%, F6\textsubscript{u}: 57.14\%), and rely more on advertisement as their source of revenue (F17\textsubscript{s}: 100.00\%; F17\textsubscript{u}: 76.92\%). 
Finally, marginal positive effects could be induced by the presence of clear trade rules (F18\textsubscript{s}: 64.71\%; F18\textsubscript{u}: 45.83\%) and the activity of moderators (F20\textsubscript{s}: 77.78\%, F20\textsubscript{u}: 58.33\%). 


\section{Discussion}
\label{sec:discussion}

Our analysis suggests that motivated cybercriminals may tend to choose markets with specific feature sets where they engage with the illicit trade of goods and that these criteria are rather stable across the forums they choose. 

\subsection{Impartiality in trade and seller verification}
\label{subsec:res_impartial}
According to our findings in Section~\ref{subsec:results_labelling}, it emerges that motivated cybercriminals prefer to conduct trade in marketplaces that have greater attention when selecting their \textit{Key Partners} (Fisher test, $p=0.032$). That implies that high-profile and motivated sellers are willing to undergo screening procedures and payment of fees to get access to segregated marketplaces to conduct trade that punishes dishonest traders. Another business aspect that seems to play a role in determining what markets are going to be the trade venues for motivated sellers is \textit{Customer Relationships} (Fisher test, $p=0.083$). Customer Relationships are mostly influenced by the impartiality of the administration when dealing with disputes. In other words, the administration is not directly involved in trade; when considering the opposite scenario, the administration may give greater visibility to their products at the expense of the competition. Whereas there is a process to earn the seller status on the platform, this could cause a conflict of interests in the administration, which could be tempted to refuse some applications to not lose their market position. Therefore, it appears that these characteristics lay the foundations for flawed marketplaces, where their operators lack the incentive to provide a fair and functional trading environment for other sellers, eventually resulting in markets not populated by prominent actors. 

\subsection{Revenue streams and admin incentives matter}
The characteristics discussed in Section~\ref{subsec:results_qualitative} seem to identify, on average, two different market categories which show some different traits.
On the one hand, \textit{successful markets} generally appear to be longstanding, with not extremely large communities, performing background checks on the sellers that want to operate on the market, and their administration is not involved with trade.
\textit{Unsuccessful markets}, on the other hand, appear to focus more on offering subscriptions to their members and their administration often offers products of various natures. 
That somewhat suggests that there are two types of markets, based on their mission and revenue strategy. On one hand, we find unsuccessful marketplaces where the administrators advertise their own products, focusing on earning as much as possible from their customers with the use of subscription services which give aesthetic perks and other miscellaneous privileges on the platform, but neglect care when selecting their partners (sellers) and keeping the entrance as unrestricted as possible. The high attractiveness of such markets and the lack of interest in prosecuting scammers makes those markets more similar to the markets for lemons; some of them eventually fail, while others still thrive. A potential explanation for their apparent success could be related to the high intake of new (inexperienced) members, and to the capability of the administrators of providing products of sufficient quality, even for free. 
Differently, administrators of successful markets aim at obtaining value by providing a neutral, and competitive market, whose revenues rely on the payment of fees from sellers and promote fairness as an incentive to increase trade volume.

\subsection{Smaller, less exposed markets tend to be more successful}
\label{subsec:res_oldforums}
The apparently irrelevant aspects identified in \textit{Key Resources} presented in Section~\ref{subsec:results_labelling}, like the presence of CAPTCHAs and the possibility to enable 2FA, seem to play a negative role in the selection process. However, these could be interpreted as the effects of other causes; looking at Table~\ref{tab:communities}, it emerges that the 9 forums featuring at least one hit (the first 9 forums of the table) have on average fewer posts and are on the underground scene for more time than their counterpart. That may indicate that the latter are more exposed and easily accessible markets (that is, well-known markets, often populated by script-kiddies) than more segregated and longstanding markets.
Considering the target population of easily accessible forums (as they can be found easily by googling `hacking forums'), these are more subjected to DDoS attacks and other abuses, rather than more segregated ones. That could be interpreted as a signal of less attractive markets for prominent cybercriminals. 

\subsection{Markets actively try to remove evidence of criminal activity}
As noted in Section~\ref{subsec:results_aliases}, we identified five markets,
four of which successful, performing content removal. We noted that the post count reported in a profile of some usernames was higher than the actual number of posts available for that user; we verified across all the forums that it is not possible to delete your own content, suggesting then that this is an action performed from the market administrators. 
Furthermore, three markets performed a shadow-banning of some profiles on their market. For example, in \crdclub, we find that five of the nine removed profiles were associated with members of the Infraud Organization~\cite{dojinfraud}, a case connected to a large-scale carding organization; another two were connected to the Kostyukov case~\cite{dojkostyukov}, where the 39 alleged cybercriminals were accused for being part of a racketeering scheme and counterfeit documents production; another user was the admin of the infamous marketplace Darkode, taken down from the police. That indicates that evidence of major criminal endeavors may disappear from related markets with time, suggesting that retrospective studies may systematically \textit{not} include major criminal service or technological solutions in their analyses.

\subsection{User expectations may signal `virtuous' market forums}
During our investigation, we noted a peculiar mechanism emerging from the discussion in the dispute resolution sections of two `successful' markets, \xss\ and \exploit. 
In July 2022, a user advertised on \xss\ a 1-click 0-day trojan for Android and iOS for an undisclosed price of `8 digit price (USD)' (sic.). The alleged software is a stolen copy of the complete source code of governmental use spyware from a cyberwarfare company. 
Whereas not specified in the forum rules, several members pointed out that to execute deals of this proportion, depositing a conspicuous amount of money is necessary as proof of commitment; a longstanding member with high status in the community reports:
\blockquote{
\textit{You seller what kind of guarantees can you offer? You don't even have a 500k/1M deposit on forum (Which is required to sell products here). [...] I honestly believe that even the most honest escrow would be tempted once he sees 50.000.000\$ in Crypto deposited in his wallet.}}
The seller brings additional evidence supporting the idea that they own the product (e.g., screenshots of the manual, a quotation from the company with the list of features included, ...), but the community remained unconvinced. 
The same seller was banned from \exploit\ because the seller and a potential buyer failed to converge on the execution modality of the transaction, raising the suspicion that the seller was acting in bad faith. In particular, the seller proposed that part of the funds would have been directly transferred to them, while the remaining part would have been transferred to an escrow service. 
This episode highlights how user expectations and community effects can make up for the lack of rules for specific edge cases. In turn, this may signal a virtuous market forum where these expectations can thrive, or at least develop.


\subsection{Is this the full picture?} To discuss the features that support successful trade in markets, we rely on the principles of moral hazard and adverse selection, and consider how market features inform the business model of an underground community. 
From our analysis, we do not see a `clear cut' distinction between the properties of successful and unsuccessful markets, and we find that, for example, less regulated markets can still thrive. We are aware that agency theory and the framing of underground marketplaces as businesses may still be insufficient to grasp the full extent of issues that characterize trade in mutually distrusted relations. Hence, we ask: \textit{how can this picture be improved?}  Several scientists from different disciplines have made efforts to study how markets can thrive from different angles; from a criminologist perspective, Soudijn and Zegers study the (private) communications among participants to a carding market obtained from a leak and discuss how the properties of a forum ease the execution of a criminal script~\cite{soudijn2012cybercrime}. In particular, they note the analogies of forums with the concept of `offender convergence setting', first formulated by Felson~\cite{felson2003process}. With this concept, Felson describes locations where criminals converge for relaxation, exchange of information, and trade. 
Lusthaus analyzes the problem of trust in cybercriminal communities from a sociological point of view~\cite{lusthaus2012trust} and identifies three major problems: establishing cybercriminal identities, assessing cybercriminal attributes, and extra-legal governance. 
Lusthaus notes that marketplaces use several mechanisms to mitigate these problems and to keep scammers and police out of their communities. Gambetta and Lusthaus note that many of those mechanisms are based on the production of signals, which can be trusted especially when they are costly to produce~\cite[Chap.1]{gambetta2011codes}, \cite{lusthaus2012trust}. 

The question remains of which additional theoretical angles and empirical measures should be accounted for to integrate (or re-invent) the current framework. We consider this part of future work and hope that the WEIS community could help us in identifying promising ways forward.

\section{Limitations}
\label{sec:limitations}

\smallskip \noindent
\textbf{Data representativeness.}
We cannot claim that our data is fully representative of existing underground market forums, or arrested cybercriminals. 
To build our ground truth, we relied on \texttt{arresttracker.com}, a website (now offline) that collects a list of cases related to cybercrime spanning from 1970 to 2021. The website is run on a voluntary basis and provides no warranties on the completeness and correctness of the collected and reported cybercriminal cases. The reported data comes from public statements from the USA Department of Justice, Europol press releases, or cybersecurity media outlets, and sources are reported for the majority of cases. Therefore, our study is limited only to criminals that have been observed by the USA authorities, including those that have been prosecuted during collaborations with international law enforcement agencies. Further, the inclusion of additional forums may provide different insights into the data. We share the dataset at \url{https://security1.win.tue.nl}.

\smallskip \noindent
\textbf{Missing `hits'.}
To minimize the risk of including false positives in our findings, we define a stringent criterion to correlate market usernames, activity, alias, accusations, and the temporal correspondence between the criminal activity in the forum and the start of the prosecution. 
As a result, we discard little variations of usernames that could be related to the activity of copycats. However, this assumes that the aliases we collected are correct and accurate, and that their users have used them consistently across all forums. For example, in \fuckav\ we identified a user registered with the alias `† Voland †' and, despite having in our list the alias `Voland', we refrain to include it in the analysis. Further, during our investigation, we noted that some accounts in some markets reported a message count greater than the actual messages that we could retrieve. That leads us to believe that said markets remove content from certain profiles as a consequence of an indictment to limit the risk of collateral damage to the community or upon account owners' request. Similarly, in some cases, we found evidence of accounts that have been renamed following an indictment. We witnessed two cases in \hackforums\ where these users wrote a `farewell letter' to their customers, explaining the reasons that lead them to end business, and warning them of the risks they may incur into. To that extent, these profiles remain hard to identify and even manual investigation may be insufficient when the content of their advertisements has been redacted or removed entirely. Finally, we do not account for the actual existence of the forums in our selection before the time of arrest of each convicted cybercriminal. That may bias our results as it may not reflect the options available to each cybercriminal on which forums to trade on before they were convicted. However, as we consider convicted cybercriminals over a period within which all forums have been active (and that all forums but two have at least an `inconclusive hit'), we consider any bias generated by this to be unlikely.

\smallskip \noindent
\textbf{Markets evolution.}
In this study, we evaluate market features in accordance with the proposed framework. We inspected the markets over the period that spans from October 2022 and February 2023. That allows us to draw considerations on the \textit{current state} of the markets in relation to the identified hits. However, this does not account for any evolution in the markets' settings, and that indicted/convicted criminals could have based their choice on a different set of features. As such, the proposed framework should be considered as an evaluation instrument about the current state of markets to believably support the trade of cybercriminal technology. In fact, a market presenting the aspects of a `successful' one does \textit{not} necessarily imply that trade of effective criminal technology must happen, but rather that it could represent a viable alternative in case other marketplaces do not convincingly establish trust among participants and in the \textit{bona fide} of their administration (anymore). 

\section{Related work}
\label{sec:relatedwork}

\subsection{Threat intelligence limitations}
Currently, the cybercriminal landscape is dotted with communities serving the purpose of encountering like-minded people to exchange information on how to perform unlawful activities and trade the products of their misconduct. From a threat intelligence (TI) perspective, the possibility of joining these communities offers valuable insights into attacker behavior and emerging threats to produce early indicators of malicious activity. 
However, in the last few years, some studies have highlighted how TI suffers from coverage and accuracy problems~\cite{thomas2016abuse, li2019reading}. Bouwman et al. empirically evaluated the threat indicators overlap for $22$ threat actors produced from two leading commercial TI vendors, finding marginal to no overlap among them~\cite{bouwman2020different}. A study reported how defenders, on average, rely on $7.7$ TI providers to ensure coverage in their threat feeds~\cite{ponemon2019value}, which is not always a viable solution due to the high fees those services impose. The adoption of TI feeds from multiple providers further exacerbates the problem of responding to real threats; the amount of Indicators of Compromise used in Intrusion Detection Systems produces an overwhelming volume of false positives, hindering the capability of performing timely actions in a Security Operation Center~\cite{rosso2022saibersoc, shah2018understanding}. The problem with the volume of generated indicators also lies in the difficulty of extracting the most relevant signals from the threat landscape, rather than the most obvious ones~\cite{bouwman2020different}. 
That suggests that there could be a problem in the assessment of what are the sources of valuable threat intelligence and that a remarkable amount of noise could be collected in the process~\cite{huang2018systematically, thomas2015framing}. 
    
\subsection{Underground ecosystem characterization}
By analogy, research faces similar problems too. There are multiple studies analyzing new markets, their population, and how they interact, but it remains unclear whether these findings capture the full picture of the underground economy~\cite{huang2018systematically, thomas2015framing}. In fact, some studies suggest how questionable the quality of certain offered goods may be in relation to their price when no market regulatory mechanism is in place~\cite{herley2010nobody, yip2013trust}. 
Previous research studying high-profile marketplaces for 0-day exploits spotlighted how these communities implement different mechanisms to establish trust among peers~\cite{allodi2017economic}. In criminal settings, where interaction happens among mutually distrusted parties, several mechanisms must be in place to mitigate the insurgence of dishonest behavior, facilitated by the information asymmetry and ultimately leading to adverse selection (i.e., buyers can not distinguish between good and bad products)~\cite{herley2010nobody, allodi2016then, soudijn2012cybercrime, yip2013trust}. Some studies highlighted how illicit marketplaces may limit the influx of new members in a community via the adoption registration fees, or vetting registrations via interviews, or using pull-in mechanisms where members of the community guarantee (`vouch') for the intentions of the new applicant~\cite{allodi2017economic, georgoulias2021qualitative, huang2018systematically}. As an effect, these restrictions severely limit the chances of security researchers infiltrating these communities~\cite{thomas2015framing}, which face the ethical concerns of (economically) supporting cybercriminal ventures. In other scenarios, constructing a credible online identity to obtain clearance is deemed too (economically and timely) expensive to pursue~\cite{schafer2019blackwidow}. As a result, some illicit marketplaces may remain out of reach. In research, studies on underground markets for which data leaks are available are not uncommon~\cite{yip2013forums, dupont2017darkode, motoyama2011analysis, overdorf2018under, afroz2013honor}, or studies including specific markets for being `famous'~\cite{georgoulias2021qualitative}, `popular'~\cite{bhalerao2019mapping}, or `large'~\cite{pastrana2019measuring, dupont2016ecology, lusthaus2012trust}. Notwithstanding the valuable contributions these studies made to our comprehension of cybercrime, it remains unclear if these selection mechanisms are shared by motivated cybercriminals too, and whether these assumptions introduce biases in our understanding of cybercrime at large. In fact, `fame' could be attributed to a large user base or abundance of content, which in turn could be a proxy of their ease of access, indicate the presence of spam/low-quality content, or even how they are ranked on search engines. If this is the case, their selection as study subjects may be representative only of specific segments of the underground economy, while more segregated markets may remain unexplored. However, closed access is not the only aspect that may be involved in creating flourishing markets fueled by high-quality products. Hence, understanding what are the features of successful markets is critical to grasp the criminal selection process when choosing the venue to conduct their criminal operations, helping researchers to make a more informed decision on the markets of choice for their research. 


\section{Conclusions}
\label{sec:conclusions}

In light of that, we acknowledge that this work cannot fully picture the complex phenomenons that characterize the cybercriminal activity of underground forums; indeed, the endogenous and exogenous phenomenons that impact the rise and fall of markets are too multifaceted to be captured from a single framework. Nonetheless, we hope that our contribution may help to improve our current understanding of the complexity of cybercrime, and will inspire further research from multiple disciplines. We believe that the answer to this complex question is yet to be found\footnote{Someone may argue that the answer has already been found, but it is unclear to which \textit{question} `$42$' is the answer.}, and the key lies in the joint efforts from multiple disciplines studying the dynamics driving this fast-paced threatscape.

\smallskip
\noindent\textbf{Acknowledgments.} 

This work is supported by the ITEA3 programme through the DEFRAUDIfy project funded by Rijksdienst voor Ondernemend Nederland, Grant No. ITEA191010, and by the INTERSCT project, Grant No. NWA.1162.18.301, funded by Netherlands Organisation for Scientific Research (NWO). We thank our industrial partner Web-IQ for providing us with valuable historical information from underground forums.


\bibliographystyle{IEEEtran}

\bibliography{references}

\begin{thebibliography}{10}
\providecommand{\url}[1]{#1}
\csname url@samestyle\endcsname
\providecommand{\newblock}{\relax}
\providecommand{\bibinfo}[2]{#2}
\providecommand{\BIBentrySTDinterwordspacing}{\spaceskip=0pt\relax}
\providecommand{\BIBentryALTinterwordstretchfactor}{4}
\providecommand{\BIBentryALTinterwordspacing}{\spaceskip=\fontdimen2\font plus
\BIBentryALTinterwordstretchfactor\fontdimen3\font minus
  \fontdimen4\font\relax}
\providecommand{\BIBforeignlanguage}[2]{{%
\expandafter\ifx\csname l@#1\endcsname\relax
\typeout{** WARNING: IEEEtran.bst: No hyphenation pattern has been}%
\typeout{** loaded for the language `#1'. Using the pattern for}%
\typeout{** the default language instead.}%
\else
\language=\csname l@#1\endcsname
\fi
#2}}
\providecommand{\BIBdecl}{\relax}
\BIBdecl

\bibitem{campobasso2020impersonation}
M.~Campobasso and L.~Allodi, ``{Impersonation-as-a-Service: Characterizing the
  Emerging Criminal Infrastructure for User Impersonation at Scale},'' in
  \emph{Proceedings of the 2020 ACM SIGSAC Conference on Computer and
  Communications Security}, 2020, pp. 1665--1680.

\bibitem{pastrana2019measuring}
S.~Pastrana, A.~Hutchings, D.~Thomas, and J.~Tapiador, ``Measuring ewhoring,''
  in \emph{Proceedings of the Internet Measurement Conference}, 2019, pp.
  463--477.

\bibitem{gambetta2011codes}
D.~Gambetta, ``{Codes of the Underworld}.''\hskip 1em plus 0.5em minus
  0.4em\relax Princeton University Press, 2011.

\bibitem{soudijn2012cybercrime}
M.~R. Soudijn and B.~C.~T. Zegers, ``Cybercrime and virtual offender
  convergence settings,'' \emph{Trends in organized crime}, vol.~15, no. 2-3,
  pp. 111--129, 2012.

\bibitem{huang2018systematically}
K.~Huang, M.~Siegel, and S.~Madnick, ``Systematically understanding the cyber
  attack business: A survey,'' \emph{ACM Computing Surveys (CSUR)}, vol.~51,
  no.~4, pp. 1--36, 2018.

\bibitem{christin2013traveling}
N.~Christin, ``{Traveling the Silk Road: A measurement analysis of a large
  anonymous online marketplace},'' in \emph{Proceedings of the 22nd
  International Conference on World Wide Web}, 2013, pp. 213--224.

\bibitem{thomas2015framing}
K.~Thomas, D.~Huang, D.~Wang, E.~Bursztein, C.~Grier, T.~J. Holt, C.~Kruegel,
  D.~McCoy, S.~Savage, and G.~Vigna, ``Framing dependencies introduced by
  underground commoditization,'' \emph{{14th Workshop on the Economics of
  Information Security (WEIS)}}, 2015.

\bibitem{alrwais2017under}
S.~Alrwais, X.~Liao, X.~Mi, P.~Wang, X.~Wang, F.~Qian, R.~Beyah, and D.~McCoy,
  ``Under the shadow of sunshine: Understanding and detecting bulletproof
  hosting on legitimate service provider networks,'' in \emph{2017 IEEE
  Symposium on Security and Privacy (SP)}.\hskip 1em plus 0.5em minus
  0.4em\relax IEEE, 2017, pp. 805--823.

\bibitem{herley2010nobody}
C.~Herley and D.~Flor{\^e}ncio, ``Nobody sells gold for the price of silver:
  Dishonesty, uncertainty and the underground economy,'' in \emph{Economics of
  Information Security and Privacy}.\hskip 1em plus 0.5em minus 0.4em\relax
  Springer, 2010, pp. 33--53.

\bibitem{akerlof}
G.~A. Akerlof, ``{The Market for "Lemons": Quality Uncertainty and the Market
  Mechanism},'' \emph{The Quarterly Journal of Economics}, vol.~84, no.~3, pp.
  488--500, 1970.

\bibitem{yip2013forums}
M.~Yip, N.~Shadbolt, and C.~Webber, ``Why forums? an empirical analysis into
  the facilitating factors of carding forums,'' in \emph{Proceedings of the 5th
  annual ACM Web Science Conference}, 2013, pp. 453--462.

\bibitem{allodi2016then}
L.~Allodi, M.~Corradin, and F.~Massacci, ``{Then and Now: On the Maturity of
  the Cybercrime Markets The Lesson That Black-Hat Marketeers Learned},''
  \emph{IEEE Transactions on Emerging Topics in Computing}, vol.~4, no.~1, pp.
  35--46, 2016.

\bibitem{dupont2016ecology}
B.~Dupont, A.-M. C{\^o}t{\'e}, C.~Savine, and D.~D{\'e}cary-H{\'e}tu, ``The
  ecology of trust among hackers,'' \emph{Global Crime}, vol.~17, no.~2, pp.
  129--151, 2016.

\bibitem{dupont2017darkode}
B.~Dupont, A.-M. C{\^o}t{\'e}, J.-I. Boutin, and J.~Fernandez, ``Darkode:
  Recruitment patterns and transactional features of ``the most dangerous
  cybercrime forum in the world'','' \emph{American Behavioral Scientist},
  vol.~61, no.~11, pp. 1219--1243, 2017.

\bibitem{motoyama2011analysis}
M.~Motoyama, D.~McCoy, K.~Levchenko, S.~Savage, and G.~M. Voelker, ``An
  analysis of underground forums,'' in \emph{Proceedings of the Internet
  Measurement Conference}, 2011, pp. 71--80.

\bibitem{yip2013trust}
M.~Yip, C.~Webber, and N.~Shadbolt, ``Trust among cybercriminals? carding
  forums, uncertainty and implications for policing,'' \emph{Policing and
  Society}, vol.~23, no.~4, pp. 516--539, 2013.

\bibitem{allodi2017economic}
L.~Allodi, ``Economic factors of vulnerability trade and exploitation,'' in
  \emph{Proceedings of the 2017 ACM SIGSAC Conference on Computer and
  Communications Security}, 2017, pp. 1483--1499.

\bibitem{gailmard2014principal}
\BIBentryALTinterwordspacing
S.~Gailmard, ``{Accountability and Principal–Agent Theory},'' in \emph{{The
  Oxford Handbook of Public Accountability}}.\hskip 1em plus 0.5em minus
  0.4em\relax Oxford University Press, 05 2014. [Online]. Available:
  \url{https://doi.org/10.1093/oxfordhb/9780199641253.013.0016}
\BIBentrySTDinterwordspacing

\bibitem{eisenhardt1989agency}
\BIBentryALTinterwordspacing
K.~M. Eisenhardt, ``Agency theory: An assessment and review,'' \emph{The
  Academy of Management Review}, vol.~14, no.~1, pp. 57--74, 1989. [Online].
  Available: \url{http://www.jstor.org/stable/258191}
\BIBentrySTDinterwordspacing

\bibitem{chohan2020opportunistic}
R.~Chohan, ``{Opportunistic Behavior in Industrial Marketing Relationships},''
  Ph.D. dissertation, Lule{\aa} University of Technology, 2020.

\bibitem{lusthaus2012trust}
J.~Lusthaus, ``Trust in the world of cybercrime,'' \emph{Global crime},
  vol.~13, no.~2, pp. 71--94, 2012.

\bibitem{pearson2003king}
G.~Pearson and D.~Hobbs, ``King pin? a case study of a middle market drug
  broker,'' \emph{The Howard Journal of Criminal Justice}, vol.~42, no.~4, pp.
  335--347, 2003.

\bibitem{goncharov2015criminal}
M.~Goncharov, ``Criminal hideouts for lease: Bulletproof hosting services,''
  \emph{Forward-Looking Threat Research (FTR) Team, A TrendLabsSM Research
  Paper}, vol.~28, 2015.

\bibitem{collier2020cybercrime}
B.~Collier, R.~Clayton, A.~Hutchings, and D.~Thomas, ``Cybercrime is (often)
  boring: maintaining the infrastructure of cybercrime economies,'' \emph{{19th
  Workshop on the Economics of Information Security (WEIS)}}, 2020.

\bibitem{collier2019booting}
B.~Collier, D.~R. Thomas, R.~Clayton, and A.~Hutchings, ``Booting the booters:
  Evaluating the effects of police interventions in the market for
  denial-of-service attacks,'' in \emph{Proceedings of the Internet Measurement
  Conference}, 2019, pp. 50--64.

\bibitem{georgoulias2021qualitative}
D.~Georgoulias, J.~M. Pedersen, M.~Falch, and E.~Vasilomanolakis, ``{A
  qualitative mapping of Darkweb marketplaces},'' in \emph{2021 APWG Symposium
  on Electronic Crime Research (eCrime)}.\hskip 1em plus 0.5em minus
  0.4em\relax IEEE, 2021, pp. 1--15.

\bibitem{franklin2007inquiry}
J.~Franklin, A.~Perrig, V.~Paxson, and S.~Savage, ``An inquiry into the nature
  and causes of the wealth of internet miscreants.'' \emph{Proceedings of the
  2007 ACM SIGSAC Conference on Computer and Communications Security}, vol.~7,
  pp. 375--388, 2007.

\bibitem{overdorf2018under}
R.~Overdorf, C.~Troncoso, R.~Greenstadt, and D.~McCoy, ``Under the underground:
  Predicting private interactions in underground forums,'' \emph{arXiv preprint
  arXiv:1805.04494}, 2018.

\bibitem{osterwalder2010business}
A.~Osterwalder and Y.~Pigneur, \emph{Business model generation: a handbook for
  visionaries, game changers, and challengers}.\hskip 1em plus 0.5em minus
  0.4em\relax John Wiley \& Sons, 2010, vol.~1.

\bibitem{bermudez2021shady}
A.~Bermudez-Villalva and G.~Stringhini, ``The shady economy: Understanding the
  difference in trading activity from underground forums in different layers of
  the web,'' in \emph{2021 APWG Symposium on Electronic Crime Research
  (eCrime)}.\hskip 1em plus 0.5em minus 0.4em\relax IEEE, 2021, pp. 1--10.

\bibitem{bhalerao2019mapping}
R.~Bhalerao, M.~Aliapoulios, I.~Shumailov, S.~Afroz, and D.~McCoy, ``Mapping
  the underground: Supervised discovery of cybercrime supply chains,'' in
  \emph{2019 APWG Symposium on Electronic Crime Research (eCrime)}.\hskip 1em
  plus 0.5em minus 0.4em\relax IEEE, 2019, pp. 1--16.

\bibitem{afroz2013honor}
S.~Afroz, V.~Garg, D.~McCoy, and R.~Greenstadt, ``Honor among thieves: A
  common's analysis of cybercrime economies,'' in \emph{2013 APWG eCrime
  Researchers Summit}.\hskip 1em plus 0.5em minus 0.4em\relax IEEE, 2013, pp.
  1--11.

\bibitem{campobasso2022threat}
M.~Campobasso and L.~Allodi, ``{THREAT/crawl: a Trainable, Highly-Reusable, and
  Extensible Automated Method and Tool to Crawl Criminal Underground Forums},''
  \emph{arXiv preprint arXiv:2212.03641}, 2022.

\bibitem{selenium}
\BIBentryALTinterwordspacing
J.~Huggins, P.~Gross, J.~T. Wang, and individual contributors, ``Selenium, a
  suite of tools for browser automation.'' 2004. [Online]. Available:
  \url{https://www.selenium.dev/}
\BIBentrySTDinterwordspacing

\bibitem{tbselenium}
\BIBentryALTinterwordspacing
G.~Acar, M.~Juarez, and individual contributors, ``{tor-browser-selenium - Tor
  Browser automation with Selenium},'' 2020. [Online]. Available:
  \url{https://github.com/webfp/tor-browser-selenium}
\BIBentrySTDinterwordspacing

\bibitem{arresttracker}
\BIBentryALTinterwordspacing
{CTRLBOX Consulting}, ``{Cyber Crime Incident Tracker - Cached copy (website
  offline)},'' 2021. [Online]. Available:
  \url{https://web.archive.org/web/20221205110731/https://www.arresttracker.com/pages}
\BIBentrySTDinterwordspacing

\bibitem{klimenko2}
\BIBentryALTinterwordspacing
{Department of Justice}, ``Text of the indictment charging 36 defendants for
  alleged roles in transnational criminal organization responsible for
  cybercrimes,'' \emph{Equipo Nizkor}, 2018. [Online]. Available:
  \url{http://www.derechos.org/nizkor/corru/doc/infraud2.html}
\BIBentrySTDinterwordspacing

\bibitem{pastrana2018crimebb}
S.~Pastrana, D.~R. Thomas, A.~Hutchings, and R.~Clayton, ``Crimebb: Enabling
  cybercrime research on underground forums at scale,'' in \emph{Proceedings of
  the 2018 World Wide Web Conference}, 2018, pp. 1845--1854.

\bibitem{martin2016ethics}
J.~Martin and N.~Christin, ``Ethics in cryptomarket research,''
  \emph{International Journal of Drug Policy}, vol.~35, pp. 84--91, 2016.

\bibitem{benjamin2019dice}
V.~Benjamin, J.~S. Valacich, and H.~Chen, ``Dice-e: A framework for conducting
  darknet identification, collection, evaluation with ethics.'' \emph{MIS
  Quarterly}, vol.~43, no.~1, 2019.

\bibitem{turk2020tight}
K.~Turk, S.~Pastrana, and B.~Collier, ``A tight scrape: methodological
  approaches to cybercrime research data collection in adversarial
  environments,'' in \emph{2020 IEEE European Symposium on Security and Privacy
  Workshops (EuroS\&PW)}.\hskip 1em plus 0.5em minus 0.4em\relax IEEE, 2020,
  pp. 428--437.

\bibitem{laferriere2023examining}
D.~Laferri{\`e}re and D.~D{\'e}cary-H{\'e}tu, ``Examining the uncharted dark
  web: Trust signalling on single vendor shops,'' \emph{Deviant Behavior},
  vol.~44, no.~1, pp. 37--56, 2023.

\bibitem{leukfeldt2015organised}
E.~R. Leukfeldt, ``Organised cybercrime and social opportunity structures: A
  proposal for future research directions,'' \emph{The European Review of
  Organised Crime}, vol.~2, no.~2, pp. 91--103, 2015.

\bibitem{tabarrok2015end}
A.~Tabarrok and T.~Cowen, ``The end of asymmetric information,'' \emph{Cato
  Unbound}, vol.~6, 2015.

\bibitem{soska2015measuring}
K.~Soska and N.~Christin, ``Measuring the longitudinal evolution of the online
  anonymous marketplace ecosystem,'' in \emph{Proceedings of the 24th USENIX
  Security Symposium}, 2015, pp. 33--48.

\bibitem{hutchings2015crime}
A.~Hutchings and T.~J. Holt, ``A crime script analysis of the online stolen
  data market,'' \emph{British Journal of Criminology}, vol.~55, no.~3, pp.
  596--614, 2015.

\bibitem{schelling1980strategy}
T.~C. Schelling, \emph{{The Strategy of Conflict: with a new Preface by the
  Author}}.\hskip 1em plus 0.5em minus 0.4em\relax Harvard university press,
  1980.

\bibitem{vx-underground_2022}
\BIBentryALTinterwordspacing
Vx-Underground, ``The staff of xss appear to be mildly frustrated with threat
  intelligence companies scraping their forum.they are now allowing companies
  the ability to scrape the forum for an annual fee of \$2,000.
  pic.twitter.com/pffkhxlnfr,'' Oct 2022. [Online]. Available:
  \url{https://twitter.com/vxunderground/status/1585368524901748736?s=20}
\BIBentrySTDinterwordspacing

\bibitem{abrams_2021}
\BIBentryALTinterwordspacing
L.~Abrams, ``Dutch police post "say no to cybercrime" warnings on hacker
  forums,'' Feb 2021. [Online]. Available:
  \url{https://www.bleepingcomputer.com/news/security/dutch-police-post-say-no-to-cybercrime-warnings-on-hacker-forums/}
\BIBentrySTDinterwordspacing

\bibitem{decary2015sifting}
D.~D{\'e}cary-H{\'e}tu and J.~Aldridge, ``Sifting through the net: Monitoring
  of online offenders by researchers,'' \emph{European Review of Organised
  Crime}, vol.~2, no.~2, pp. 122--141, 2015.

\bibitem{decary2016criminals}
D.~D{\'e}cary-H{\'e}tu and A.~Lepp{\"a}nen, ``Criminals and signals: An
  assessment of criminal performance in the carding underworld,''
  \emph{Security Journal}, vol.~29, pp. 442--460, 2016.

\bibitem{santanna2015booters}
J.~J. Santanna, R.~van Rijswijk-Deij, R.~Hofstede, A.~Sperotto, M.~Wierbosch,
  L.~Z. Granville, and A.~Pras, ``{Booters—An analysis of DDoS-as-a-service
  attacks},'' in \emph{2015 IFIP/IEEE International Symposium on Integrated
  Network Management (IM)}.\hskip 1em plus 0.5em minus 0.4em\relax IEEE, 2015,
  pp. 243--251.

\bibitem{karami2013understanding}
M.~Karami and D.~McCoy, ``{Understanding the emerging threat of
  DDoS-as-a-Service},'' in \emph{6th USENIX Workshop on Large-Scale Exploits
  and Emergent Threats (LEET 13)}, 2013.

\bibitem{dojinfraud}
\BIBentryALTinterwordspacing
{Department of Justice}, ``{First Superseeding Indictment}.'' [Online].
  Available:
  \url{https://web.archive.org/web/20221029212509/https://dd80b675424c132b90b3-e48385e382d2e5d17821a5e1d8e4c86b.ssl.cf1.rackcdn.com/external/infraudsupersedingindictment.pdf}
\BIBentrySTDinterwordspacing

\bibitem{dojkostyukov}
\BIBentryALTinterwordspacing
{Department of Justice }, ``{Criminal Indictment}.'' [Online]. Available:
  \url{https://web.archive.org/web/20150211203259/https://cis.uab.edu/forensics/blog/Kostyukov.Indictment.pdf}
\BIBentrySTDinterwordspacing

\bibitem{felson2003process}
M.~Felson, ``The process of co-offending,'' \emph{Crime prevention studies},
  vol.~16, pp. 149--168, 2003.

\bibitem{thomas2016abuse}
K.~Thomas, R.~Amira, A.~Ben-Yoash, O.~Folger, A.~Hardon, A.~Berger,
  E.~Bursztein, and M.~Bailey, ``The abuse sharing economy: Understanding the
  limits of threat exchanges,'' in \emph{International Symposium on Research in
  Attacks, Intrusions, and Defenses (RAID)}.\hskip 1em plus 0.5em minus
  0.4em\relax Springer, 2016, pp. 143--164.

\bibitem{li2019reading}
V.~G. Li, M.~Dunn, P.~Pearce, D.~McCoy, G.~M. Voelker, S.~Savage, and
  K.~Levchenko, ``Reading the tea leaves: A comparative analysis of threat
  intelligence,'' in \emph{{Proceedings of the 28th USENIX Security
  Symposium}}, 2019.

\bibitem{bouwman2020different}
X.~Bouwman, H.~Griffioen, J.~Egbers, C.~Doerr, B.~Klievink, and M.~Van~Eeten,
  ``A different cup of $\{$TI$\}$? the added value of commercial threat
  intelligence,'' in \emph{{Proceedings of the 29th USENIX Security
  Symposium}}, 2020, pp. 433--450.

\bibitem{ponemon2019value}
\BIBentryALTinterwordspacing
{Ponemon Institute LLC}, ``The value of threat intelligence: Annual study of
  north american \& united kingdom companies.'' [Online]. Available:
  \url{https://stratejm.com/wp-content/uploads/2019/08/2019_Ponemon_Institute-Value_of_Threat_Intelligence_Research_Report_from_Anomali.pdf}
\BIBentrySTDinterwordspacing

\bibitem{rosso2022saibersoc}
M.~Rosso, M.~Campobasso, G.~Gankhuyag, and L.~Allodi, ``{SAIBERSOC: A
  Methodology and Tool for Experimenting with Security Operation Centers},''
  \emph{Digital Threats: Research and Practice (DTRAP)}, vol.~3, no.~2, pp.
  1--29, 2022.

\bibitem{shah2018understanding}
A.~Shah, R.~Ganesan, S.~Jajodia, and H.~Cam, ``{Understanding tradeoffs between
  throughput, quality, and cost of alert analysis in a CSOC},'' \emph{IEEE
  Transactions on Information Forensics and Security}, vol.~14, no.~5, pp.
  1155--1170, 2018.

\bibitem{schafer2019blackwidow}
M.~Sch{\"a}fer, M.~Fuchs, M.~Strohmeier, M.~Engel, M.~Liechti, and V.~Lenders,
  ``Blackwidow: Monitoring the dark web for cyber security information,'' in
  \emph{2019 11th International Conference on Cyber Conflict (CyCon)}, vol.
  900.\hskip 1em plus 0.5em minus 0.4em\relax IEEE, 2019, pp. 1--21.

\end{thebibliography}

\appendix

\section{Market mechanisms}

In this section, we enumerate and briefly describe the specific implementations we identified in the markets under analysis, for each of the mechanisms outlined earlier in this section. Unless otherwise specified, the features are considered binary.

\smallskip
\noindent\textbf{Restricted access model.}
We find several levels of access restriction in the cybercrime forums under analysis:

\feature{F1.} \textit{Restricted sections - pull-in.} Access to one or more sections of the market can be granted by a pull-in mechanism (administrator authorization, application, community vote, invite from member).

\feature{F2.} \textit{Restricted sections - payment.} Access to one or more sections of the market can be granted upon the payment of a recurrent or one-off fee.

\feature{F3.} \textit{Restricted registration - pull-in.} Access to the entire market can be granted by a pull-in mechanism (administrator authorization, application, community vote, invite from member).

\feature{F4.} \textit{Restricted registration - payment.} Access to the entire market that could be granted upon the payment of a recurrent or one-off fee.

\smallskip
\noindent\textbf{Dispute resolution system.}
Markets in our selection implement dispute resolution systems in the following ways:

\feature{F5.} \textit{Scammers banned.} We find evidence that scammers are be banned from the market when found guilty.

\feature{F6.} \textit{Working dispute resolution system.} We find evidence that the market offers an active dispute resolution system where users and administrators interact on the basis of provided evidence to evaluate each case.

\feature{F7.} \textit{Neutral mediator.} We do not find any evidence that the dispute mediator(s) have other functions in the market or act as seller themselves.

\smallskip
\noindent\textbf{Reputation systems.}
Reputation systems in our market selection are characterized by:

\feature{F8.} \textit{Seller status - verification.} The market allows sellers to obtain a privileged `seller status' via manual verification.

\feature{F9.} \textit{Seller status - payment.} The market requires sellers to pay a fee to obtain a privileged `seller status'.

\feature{F10.} \textit{User status - verification.} The market allows any users to obtain a generic privileged status via manual verification.

\feature{F11.} \textit{User status - payment.} The market requires any users to pay a fee to obtain a generic privileged status.

\feature{F12.} \textit{Reputation change on trade.} A reputation score can exclusively be changed as a consequence of trading activities, as opposed to arbitrarily at any time and by any user.

\feature{F13.} \textit{Reputation change by VIP.} A reputation score can exclusively be changed by members with a high status on the market, as opposed to by any user in the forum.

\smallskip
\noindent\textbf{Mitigation of perverse incentives.}
Forum markets in our selection implement several features that may mitigate  perverse incentives leading to, for example, exit scams by the market administrators.

\feature{F14.} \textit{Seller status - recurrent fee.} By payment of a periodic fee to the market administrators, sellers can maintain a seller privileged status. 

\feature{F15.} \textit{User status - recurrent fee.} By payment of a periodic fee to the market administrators, users can maintain a generic privileged status. 

\feature{F16.} \textit{Escrow fee.} The market imposes an escrow fee imposed to all transactions using the escrow service. Fees can be a fixed amount in fiat currency, and a fixed or variable percentage of the transaction value. This feature is excluded from the analysis.

\feature{F17.} \textit{Sponsored ads.} The market offers the possibility to pay for sponsored ads.

\smallskip
\noindent\textbf{Rule system.}
To provide a common baseline for trade, there should be rules and these should be enforced. Within our markets, we find the following:
\feature{F18.} \textit{Clear trade rules.} Trade activities are regulated by a dedicated and enforceable set of rules.

\feature{F19.} \textit{Moderator roles.} The role of moderators in the market is defined by specific rules in the market. We define a non-binary scale to classify the possible outcomes:
\begin{enumerate}
    \setcounter{enumi}{-1}
    \item no moderation
    \item moderators exist
    \item moderators exist and can be looked up from a member list
    \item moderators exist and operate in specific sections 
    \item moderators exist and rules explicitly mention their roles
    \item combines the properties of 2 and 4
    \item combines the properties of 3 and 4
    \item combines the properties of 2, 3 and 4.
\end{enumerate} 
This feature is excluded from the analysis.

\feature{F20.} \textit{Active moderation.} Moderators appear to be active on the market (e.g.,conducting moderating activities during trials).

\smallskip
\noindent\textbf{Strong authentication and anti-bot features}
To limit unwanted activity and verify the human nature of users, markets in our selection implement the following:

\feature{F21.} \textit{2FA available.} The market supports two-factor-authentication for its users.

\feature{F22.} \textit{CAPTCHA on authentication.} The authentication procedure to the market is protected by CAPTCHAs.

\feature{F23.} \textit{CAPTCHA on access.} CAPTCHAs are prompted at each market access.

\smallskip
\noindent\textbf{Escrow system.}
Escrowing systems in our market selection can be:

\feature{F24.} \textit{Escrow available.} The market provides its members with an escrow system for transactions.

\feature{F25.} \textit{Escrow recommended.} The market explicitly recommends its members to use an escrow system for transactions.

\smallskip
\noindent\textbf{Interaction model.}
Interaction among participants in our selection of forum markets can be:

\feature{F26.} \textit{Public interaction.} The market allows users to publicly interact with all members.

\feature{F27.} \textit{Private interaction.} The market allows users to privately interact with all members.

\feature{F28.} \textit{Pay to show content.} Authors of a post or comment can require other users to pay a fee to show posted content.

\vspace{12pt}
\end{document}